\documentclass[letterpaper]{article}

\usepackage[T1]{fontenc}

\usepackage{geometry}
\usepackage{upgreek}
\usepackage{setspace}
\usepackage{booktabs}
\usepackage{graphicx}
\usepackage{float}
\newfloat{scheme}{htbp}{los}
\floatname{scheme}{Scheme}
\floatname{chart}{Chart}
\newfloat{graph}{htbp}{loh}

\usepackage{chemformula} % Formulas using \ch{}
\usepackage[version = 4]{mhchem} % Formulas using \ce{}

\newcommand*\mycommand[1]{\texttt{\emph{#1}}}

\usepackage{authblk}
\author[1]{Heena \dag}
\author[2]{Vineet Kumar Pandey \dag}
\author[3]{Ambesh Dixit}
\author[1]{Anver Aziz}
\author[4]{Sung Gu Kang}
\author[5]{K. C. Bhamu*}
\affil[1]{Department of Physics, Jamia Millia Islamia, New Delhi 110025, India}
\affil[2]{Department of Physics, Govt Naveen College Janakpur M.C.B. 497778, Chhatisgarh, INDIA}
\affil[3]{Advanced Materials and Device Laboratory, Department of Physics, Indian Institute of Technology, Jodhpur 342037, India}
\affil[4]{School of Chemical Engineering, University of Ulsan, Ulsan, Daehak-ro 93, Nam-gu, Ulsan 44610, Republic of Korea}
\affil[5]{Department of Physics, SLAS, Mody University of Science and Technology, Lakshmangarh, Sikar, Rajasthan, 332311, INDIA}

\title{Distorted polyhedral architecture enabled high thermoelectric performance of columnar double halide perovskites Cs$_2$AgPdCl$_5$ and Cs$_2$AgPtCl$_5$}
\date{*Email: kcbhamu85@gmail.com, \dag : These authors contributed equally}

\begin{document}

\maketitle

\begin{abstract}
  We investigate the thermoelectric properties of two newly synthesized columnar double halide perovskites Cs$_2$AgPdCl$_5$ and Cs$_2$AgPtCl$_5$. These materials accommodate a distorted local polyhedral architecture with tetrahedral symmetry compared to traditional double halide perovskites. By employing density functional theory along with the semiclassical transport model, we have analyzed the electronic and transport properties of these materials. Our results show that at 800 K, the largest figure of merit (zT) is 1.30(0.86) for p-type (n-type) Cs$_2$AgPdCl$_5$ and 0.87 for n-type Cs$_2$AgPtCl$_5$ at doping concentration $1.94 \times 10^{20}$ ($3.76 \times 10^{19}$) cm$^{-3}$ and $3.52 \times 10^{19}$ cm$^{-3}$, respectively. Remarkably, a very low doping concentration is required to achieve a high zT, setting these materials apart from others in this field. Our calculations demonstrate that Cs$_2$AgPdCl$_5$ benefits from the presence of conduction and valence band valleys near the band edges, however, the flat bands present in the valence band of Cs$_2$AgPtCl$_5$ do not improve its thermoelectric performance. Among these systems, the hole doping in Cs$_2$AgPdCl$_5$ has shown remarkable thermoelectric performance. Interestingly, the local octahedral distortions present in these perovskites contribute to a marked reduction in the lattice thermal conductivity to be 0.27 W/mK in Cs$_2$AgPtCl$_5$ and 0.20 W/mK in Cs$_2$AgPdCl$_5$ by causing enhanced scattering of phonons, further improving the thermoelectric figure of merit. This drop in thermal conductivity, combined with the favourable electronic properties, underscores the potential use of these materials for applications in highly efficient thermoelectric devices.
\end{abstract}

\section*{Keywords}
Double columar halide perovskite
Distorted octahedra
Ultralow lattice thermal conductivity
High power factor and ZT
\iffalse
\section*{Abbreviations}

Some journals require a list of abbreviations: these normally should be given
immediately after the keyswords (if required)

\fi
\section{Introduction}
The most fundamental forms of energy, such as heat and light, have attracted scientists' attention for centuries \cite{energy, energy2}. The constant supply of these resources has intrigued human minds for their use in fulfilling our energy demand. Conventional energies are depleting year by year and increasing CO$_2$ emission and pollution, which is affecting the climate adversely \cite{conv, co2}. To address these problems, Scientists have focused on enhancing the efficient use of non-conventional energy resources. Thermoelectrics is one of the ways forward for converting waste heat into clean electrical energy \cite{crc}. Thermoelectric materials have garnered a significant amount of focus in the last few decades as these materials offer promising solutions for sustainable energy harvesting and efficient refrigeration systems \cite{harvest, crc2}. In addition, thermoelectric materials are being deployed in space missions, semiconductor lasers, microchips, and unmanned missions \cite{microchip, space, unman}. The ability of materials for thermoelectric applications is governed by their figure of merit (zT) value, which is calculated as $\textrm{zT}=\frac{\sigma \textrm{S}^2 \textrm{T}}{\kappa}$, where $\sigma$, $\kappa$, S, and T denote the electrical conductivity, total thermal conductivity, Seebeck coefficient, and the absolute temperature, respectively. The efficiency of these materials can be amplified either by boosting their power factor (S$^2 \sigma$) or diminishing their thermal conductivity ($\kappa$). However, due to the intricate relationship among $\sigma$, S, and $\kappa$, the improvement of zT is not straightforward. Scientists have explored various approaches to maximize the thermoelectric power factor and minimize the thermal conductivity. For instance, the power factor can be augmented by employing resonant doping \cite{reso, reso2}, band engineering \cite{band1, band2, band3}, and carrier optimization \cite{car1, car2, car3, nbcosb}, whereas the thermal conductivity can be diminished by introducing defects \cite{nbcosb, defect1, defect2}, nanostructuring \cite{nano1, nano2, nano3}, intrinsic anharmonicity \cite{snse, bicuseo}. Thermoelectric materials such as PbTe \cite{pbte} and Bi$_2$Te$_3$ \cite{bi2te3} have shown great deployment at the commercial level in energy conversion owing to their great transport properties. Since these heavy metals are toxic and scarce in nature, the discovery of durable, innocuous, non-corrosive, and abundant element based materials is under scrutiny for efficient thermoelectric materials. One such class is perovskite materials, providing a unique combination of crystal structure such as ABX$_3$ \cite{abx3}, A$_2$BX$_6$ \cite{a2bx6}, A$_2$BB$'$X$_6$ \cite{a2bbx6}, etc (A=large cation (typically alkali or alkaline metal); B (B$'$)= monovalent (trivalent) cation (typically transition elements); X=Halogen group elements). Perovskite materials with their intriguing electronic and optical properties are playing at the forefront. Perovskite materials have been exploited intensively for research such as solar cell application \cite{solar}, LEDs and lasers \cite{leds, laser}, optoelectronics \cite{opto1, opto2}, energy storage \cite{energy1, energy2}, and thermoelectric devices \cite{thm}.  \\

\noindent Recently, inorganic/organic lead halide perovskites have been focused upon for their remarkable optical and electronic properties\cite{7} due to their high defect tolerance\cite{a5}, high absorption coefficient\cite{a3}, high scattering rate and carrier diffusion length\cite{a4}, suitable band gap and smaller exciton binding energy \cite{a6,a7}. Lately, the power conversion ability for such perovskite based solar cells has significantly reached over 25-34 $\%$ \cite{8, solar1,solar2}. A good amount of work has been reported on the lead based perovskite materials, which have shown remarkable improvement in thermoelectric figure of merit \cite{pb1,pb2,pb3,pb4,pb5,pb6}. However, lead based perovskite materials have not been commercialized at a large scale owing to the use of heavy metal and the material facing long-term instability issues in the open air \cite{instable}. Computational reports reveal that high electronic dimensionality and symmetry are the main attributions behind the excellent performance of lead halide perovskite \cite{12,13,14}. Hence, replacement of lead in such perovskite using other elements was pursued in a way that maintains its highly symmetric 3D structure \cite{leadfree}. Recently, the substitution of homovalent atoms such as Sn or Ge at Pb place has attracted considerable attention as it preserves the three dimensional perovskite structure and in turn, enables the high electronic dimensionality \cite{9,10}. However, these Sn or Ge based halide perovskites are known for capturing moisture and suffer from stability issues in the air as these are easily oxidized \cite{15}.  \\

\noindent Apart from this, an alternative approach was followed to replace Pb with a mix of univalent and trivalent cations to result in the formation of three dimensional halide double perovskite (HDP) A$_2$BB$'$X$_6$, where B represents univalent cation atoms and B$'$ denotes trivalent cation atoms, respectively \cite{10,15,16,17,18}. Such structures have the rock-salt type of arrangement with isolated BX$_6$ and B$'$X$_6$ octahedra \cite{19}. Vast combinatorial possibilities of these cations have attracted significant research, however, most of such combinations have resulted in rock salt electronic dimensionality owing to their symmetry mismatch (Cs$_2$AgBiBr$_6$) or energy mismatch (Cs$_2$NaSbCl$_6$) between B and B$'$ orbitals \cite{15,16,20,12,21}. Attributing to this rock salt electronic architecture, most of these HDPs have exhibited localized band dispersions, giving rise to heavy carriers and wide band gaps that make them unworthy of charge separation, visible light absorption, and transport properties \cite{22}. While there are few mixes of B (In and Tl) and B$'$(Sb and Bi) which have mimicked the electronic dimensionality of APbX$_3$ \cite{17,18}, these combinations could not resolve the issues associated with lead halide perovskites as In ion is unstable against oxidation and Tl is more toxic than Pb \cite{22}.    \\

\noindent The electronic properties of such materials are governed mainly by the cation at the B site along with its hybridization with the neighbouring X ions. Hence, most of the research is focused on alloying or doping at the B site for suitable modification in the electronic properties\cite{18,17,24,23}. Interestingly, the B site cation's arrangement is an additional strategy to modulating the electronic properties \cite{19}. As of now, HDPs reported so far have only rock-salt type of arrangement which is attributed to the substantial charge differences among cations at the B site \cite{25,19}. Other than these conventional structural arrangements, the B site cations can accommodate themselves in columnar (1D) or layered structures (2D). Such structural order in these HDPs exhibits 1D or 2D electronic dimensionality, respectively \cite{25,19}. With the anticipation that these HDPs can suitably modify the electronic properties, Cs$_2$AgPdCl$_5$ and Cs$_2$AgPtCl$_5$ were successfully synthesized by Ji et al. \cite{22}. These compounds were found to be indirect semiconductors with band gaps in the range of 1.33-1.77 eV \cite{22, bandgap}. These compounds have shown remarkable stability against decomposition and temperature \cite{22}. Having such interesting electronic properties and better thermal stability makes these materials eligible for further study for thermoelectric applications.
In this work, we examine the thermoelectric performance of Cs$_2$AgPdCl$_5$ and Cs$_2$AgPtCl$_5$ using density functional theory (DFT)\cite{dft} and semiclassical Boltzmann transport theory \cite{bt}. \\
\noindent The remainder of the paper is discussed as follows: section \ref{cmp} explains the computational techniques used in our calculation. In the subsequent sections, we discuss the crystal structure along with the electronic, mechanical, vibrational, and thermoelectric properties of the system. In the end, we summarize and conclude the work in section \ref{conc}. \\

\section{Computational details}
\label{cmp} 
\noindent We performed structural optimizations and total energy calculations using density functional theory (DFT) through the Vienna Ab-initio Simulation package (VASP) \cite{vasp}. We employed the projected augmented wave (PAW) pseudopotentials and the Perdew-Burke-Ernzerhof (PBE) generalized gradient approximation \cite{pbe} to model the exchange-correlation effects between electrons. The valence electron configurations for the elements Cs, Ag, Pd, Pt, and Cl were $6s^1$, $4d^{10}5s^1$, $4d^95s^1$, $5d^96s^1$, and $3s^23p^5$ respectively. During the structural relaxations, the atomic coordinates were adjusted till the residual forces fell below 0.01 eV/\AA, with an energy convergence criterion of 10$^{\text{-}6}$ eV. Appropriate k-point meshes and kinetic energy cutoffs were used for the structural optimizations ($14\times14\times20$ k-mesh, 450 eV). We computed the electronic band gap and density of states using the HSE06 hybrid functional along with the spin orbit coupling \cite{hse} with $7\times7\times10$ k-mesh. Additional analyses were carried out using supplementary codes SUMO for effective mass calculations \cite{sumo}. We calculated crystal orbital Hamilton population (COHP) using LOBSTER \cite{lobster} with k-mesh $7\times7\times10$ to reveal the bonding nature in both the systems Cs$_2$AgPdCl$_5$ and Cs$_2$AgPtCl$_5$. We used $7\times7\times10$ k-mesh to calculate elastic tensors. \\
\noindent We employed AMSET code to compute the electronic transport properties including carrier scattering rates \cite{amset}. We used the interpolation factor of 15 along with interpolation k-mesh 31 $\times$ 31 $\times$ 45 to compute the transport properties. The convergence of transport properties with respect to interpolation factor is given in supplementary information (SI), Figure SI1 for Cs$_2$AgPdCl$_5$ and Figure SI2 for Cs$_2$AgPtCl$_5$. We used the PBE+SOC computed band structure to calculate the transport properties and have utilised the HSE06+SOC computed band gaps by applying scissor correction in transport properties calculation. The primary data used for transport properties calculation is included in SI. 
\noindent We obtained the lattice thermal conductivity by solving the Boltzmann transport equation for phonons combined with the interatomic force constants of both the systems Cs$_2$AgPdCl$_5$ and Cs$_2$AgPtCl$_5$, using the ShengBTE software package\cite{sheng}. The lattice thermal conductivity is evaluated as :
\begin{equation}
\kappa_l^{\upalpha\upbeta} = \sum_{qs}C_\nu(qs)\tau(qs)v^{\upalpha}(qs)v^{\upbeta}(qs) 
\label{kl}
\end{equation}
here $\kappa_l^{\upalpha\upbeta}$ denotes the lattice thermal conductivity in $\upalpha$ and $\upbeta$ directions. The term $C_\nu(qs)$ indicates the mode-dependent heat capacity along the phonon mode of wave vector $q$ associated with branch index $s$. $\tau$ denotes the scattering rate of phonons and the component $v^{\upalpha}$ shows the phonon group velocity along $\upalpha$ direction. 

The harmonic (second-order) interatomic force constants (IFCs) are determined by the Vienna Ab-initio Simulation package (VASP) \cite{vasp} combined with the density functional perturbation theory (DFPT) with qmesh $2\times2\times3$ and kinetic energy cutoff of 450 eV. The anharmonic (third-order) IFCs for $2\times2\times3$ supercell are determined from the thirdorder.py script \cite{sheng} combined with the SCF calculations within GGA-PBE approximations. To obtain these third-order IFCs, up to 7th nearest neighbor interaction was included in the calculation. A $11\times11\times20$ q-grid was used for both the systems Cs$_2$AgPdCl$_5$ and Cs$_2$AgPtCl$_5$ to get convergence of lattice thermal conductivity calculations. We have provided the convergence of lattice thermal conductivity with respect to q grid at 300 K, Figure SI3. We computed the eigen modes at $\Gamma$ point for both the systems using Quantum ESPRESSO and DFPT \cite{qe}. We conducted the AIMD simulation in the NVT ensemble using the Nose-Hoover thermostat to confirm the structural stability at high temperature, Figure SI4.

\section{Results and discussion}
\subsection{Crystal structure}
\label{strr}
\begin{figure}[H]
\centering
  \includegraphics[height=5cm]{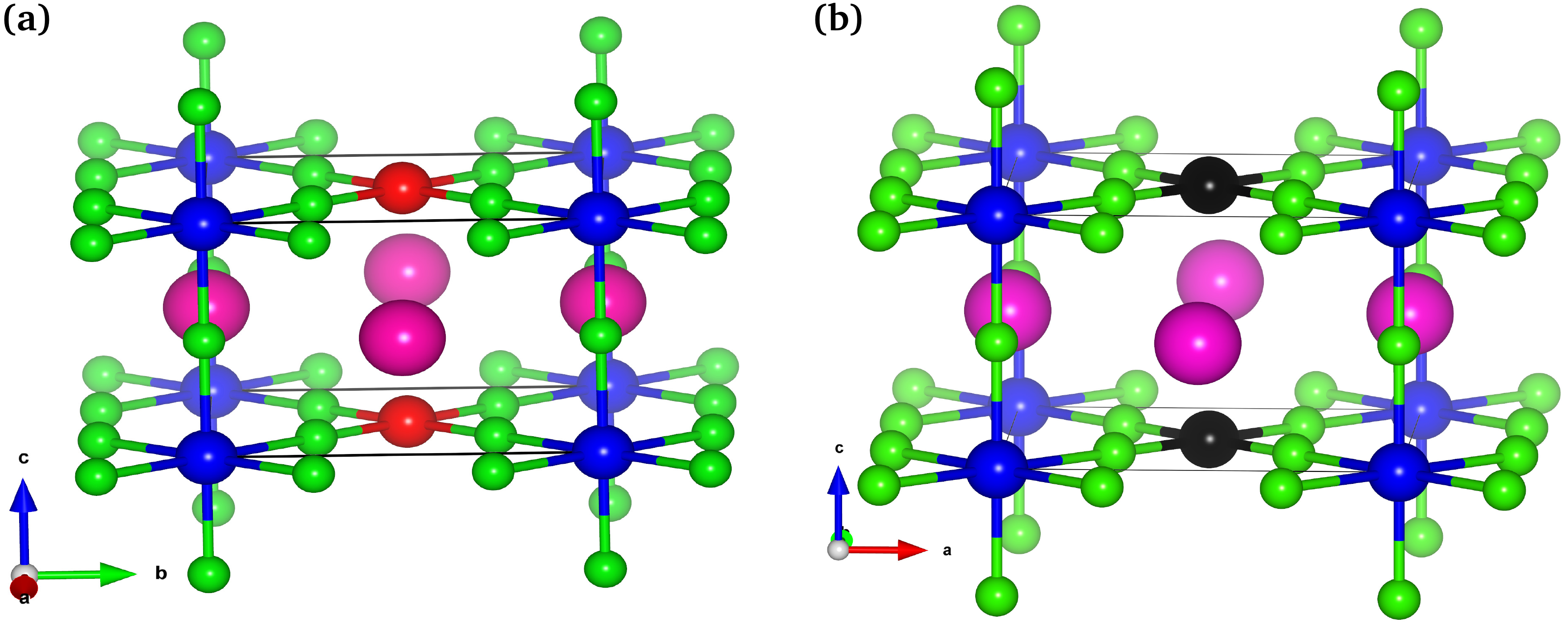}
  \caption{Crystal structure of (a) Cs$_2$AgPdCl$_5$ and (b) Cs$_2$AgPtCl$_5$. Here, red, black, blue, green, and magenta color shows Pd, Pt, Ag, Cl, and Cs atoms, respectively.}
\label{crystals}
\end{figure}

Cs$_2$AgPdCl$_5$ and Cs$_2$AgPtCl$_5$ are double columnar perovskites that crystallize in the tetragonal phase, belonging to the P4/mmm space group, as shown in Figure \ref{crystals}(a) and (b), respectively. In this configuration, both Ag and Pd/Pt atoms occupy alternating positions along the a-axis. Each Ag atom forms a distorted octahedron, coordinated with chlorine atoms, resulting in the [AgCl$_6$] structure. In contrast, the Pd/Pt atoms adopt a square planar geometry, forming [PdCl$_4$] or [PtCl$_4$], rather than the typical octahedral coordination. This difference in coordination around the Pd/Pt atoms distinguishes the double columnar perovskite structure from that of double halide perovskites. The [AgCl$_6$] octahedra share corners within the unit cell, and the large Cs atoms occupy the voids between these octahedra, balancing the overall charge. The presence of distorted [AgCl$_6$] octahedra introduces significant lattice anharmonicity in both systems as the Ag-Cl bonds are unequal, weaker and less directional as compared to Pd-Cl bonds. The calculated lattice parameters for Cs$_2$AgPdCl$_5$ (Cs$_2$AgPtCl$_5$) along the a- and c-axes are 7.31 \AA \space (7.38 \AA \space) and 4.97 \AA \space (4.95 \AA \space), respectively, which closely match previously reported values in the literature \cite{8str}. Table \ref{tabb1} compares the bond distances for both Cs$_2$AgPdCl$_5$ and Cs$_2$AgPtCl$_5$.

\begin{table}[h]

 \caption{Calculated Ag-Cl bond distance($d_{Ag\text{-}Cl}$), Cl-Pd/Pt bond distance($d_{Cl\text{-}Pd/Pt}$) and Cs-Pd/Pt bond distance($d_{Cs\text{-}Pd/Pt}$) for Cs$_2$AgPdCl$_5$ and Cs$_2$AgPtCl$_5$. All the distances are in \AA  \space unit}
\label{tabb1} 
\centering
 \begin{tabular*}{\textwidth}{@{\extracolsep{\fill}}lllll}
%\begin{tabular}{ c c c c } 
 \hline
Systems & $d_{Ag\text{-}Cl}$ & $d_{Cl\text{-}Pd/Pt}$  & $d_{Cs\text{-}Pd/Pt}$  \\
 \hline 
  Cs$_2$AgPdCl$_5$ & 2.98, 2.49 & 2.34 & 3.66 \\
  
   Cs$_2$AgPtCl$_5$  & 3.01, 2.49 & 2.33 & 4.53 \\
   \hline
\end{tabular*} 
\end{table}

\begin{figure}[H]
 \centering
 \includegraphics[height=7cm]{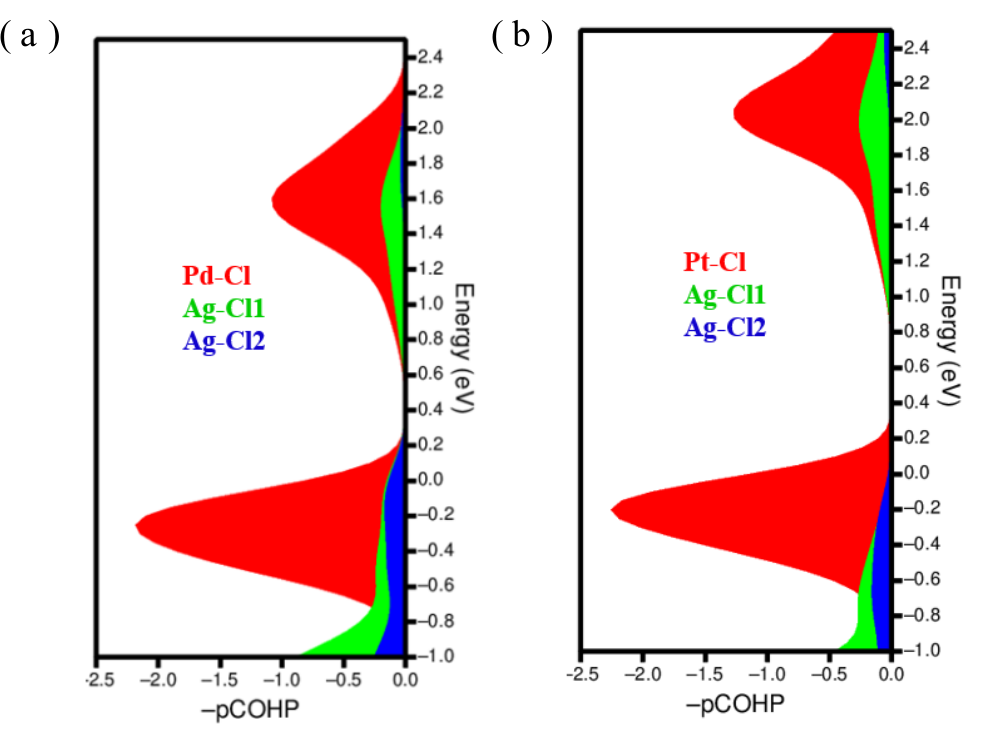}
 \caption{Projected -COHP(-pCOHP) curve for Ag-Cl1, Ag-Cl2, and Pd-Cl for (a) Cs$_2$AgPdCl$_5$ and (b) Cs$_2$AgPtCl$_5$.}
\label{cohp}
\end{figure}

\begin{figure}[h]
 \centering
 \includegraphics[height=8cm]{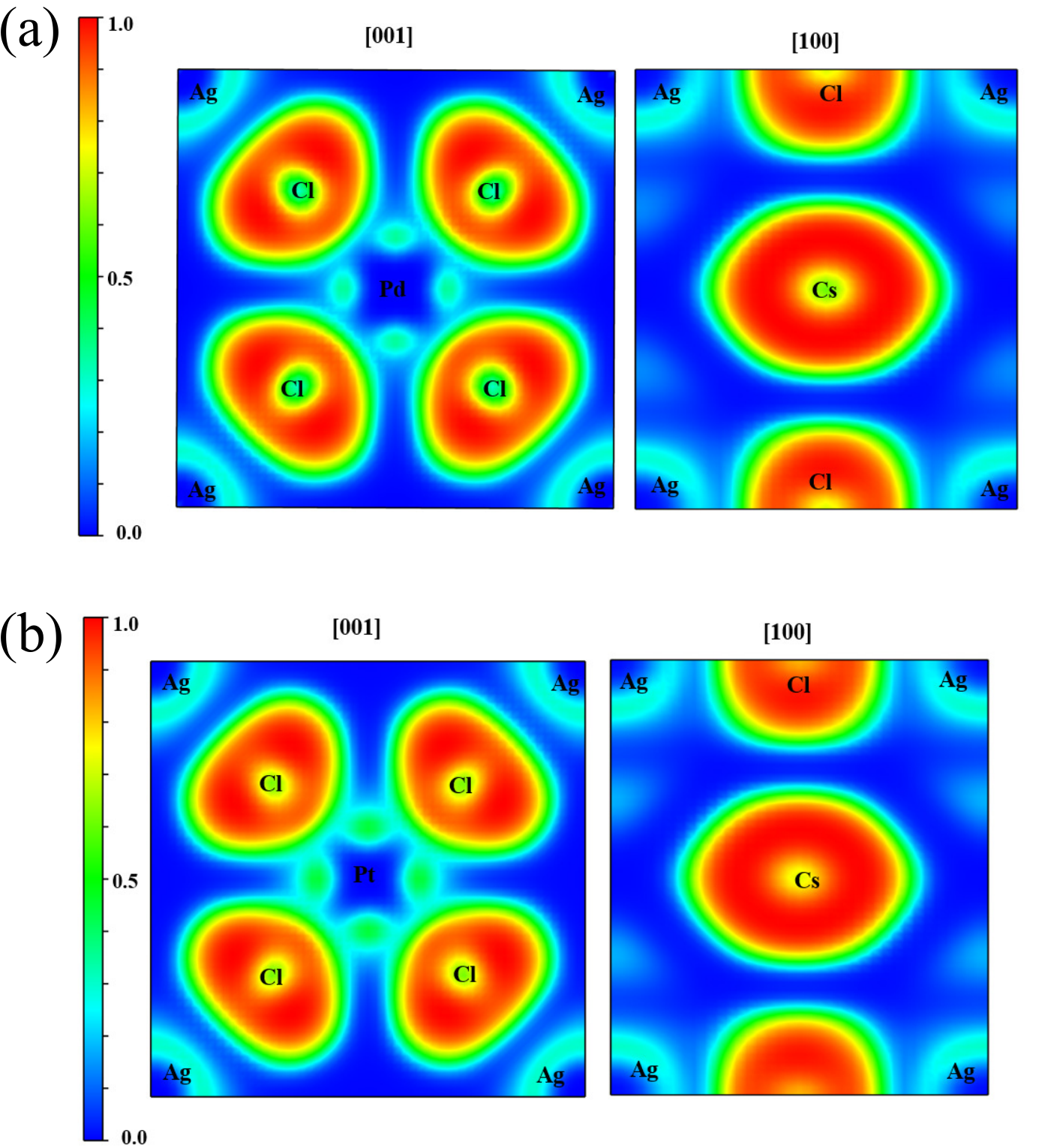}
 \caption{Electron localisation function map for (a) Cs$_2$AgPdCl$_5$ and (b) Cs$_2$AgPtCl$_5$.}
\label{elf}
\end{figure}

\noindent \paragraph*{Bonding analysis}
Figures \ref{cohp}(a) and (b) display the projected crystal orbital Hamilton population (pCOHP) curves for the Ag-Cl1, Ag-Cl2, and Pd-Cl bonds in both Cs$_2$AgPdCl$_5$ and Cs$_2$AgPtCl$_5$ systems. Here, Ag-Cl1 and Ag-Cl2 correspond to the Ag-Cl-Ag chain and Ag-Cl-Pd chain, respectively. The negative values on the X-axis represent antibonding states, with those below the Fermi energy indicating instability in the system. In our analysis, we observe higher and broader peaks for the Pd-Cl and Pt-Cl bonds just below the Fermi energy, suggesting strong antibonding interactions in both systems, with some contribution from Ag-Cl1 and Ag-Cl2 antibonding characters. A comparison of the -pCOHP peaks for both systems reveals that the Ag-Cl1 antibonding character is absent in Cs$_2$AgPtCl$_5$ at the Fermi energy, whereas it is present in Cs$_2$AgPdCl$_5$, as described in the DOS study. Additionally, the -pCOHP peak for the Pt-Cl bond is slightly narrower than that for the Pd-Cl bond, indicating a stronger interaction between the Pt 5d orbital and the Cl 3p orbital compared to the localized Pd 4d orbital and Cl 3p orbital. This is due to the smaller energy disparity between the Pt 5d and Cl 3p orbitals, with Pt 5d being more delocalized. The antibonding states in both systems extend up to 2.0 eV below the Fermi energy. The pCOHP plot for the full energy range is provided in Figure SI5. In addition to this, we have carefully analysed the electron localisation function map along [100] and [001] directions for Cs$_2$AgPdCl$_5$ and Cs$_2$AgPtCl$_5$  in the Figures \ref{elf}(a) and (b), respectively. The asymmetric lobe present at Cl atoms are oriented to Pd/Pt atom suggests strong polar covalent bond. The Ag-Cl bond is ionic in nature with delocalised electrons present in its surroundings (ELF<0.3), suggesting ionic character of Ag-Cl bond. Presence of large blue sphere around Cs suggests that Cs atom is purely ionic and does not involve in directional bonding and hence does not participate in electron conduction. In short, the Pd-Cl or Pt-Cl is predominantly involved in the chemical bonding and would participate in electronic properties. The Cs-Cl and Ag-Cl both stabilize the structure and the Ag-Cl can slightly polarize the Cl ions as the Ag has $d^{10}$ configuration. The charge density plot for both systems are given in Figure SI6.

\subsection{Electronic properties}
\begin{figure}[H]
 \centering
 \includegraphics[height=6cm]{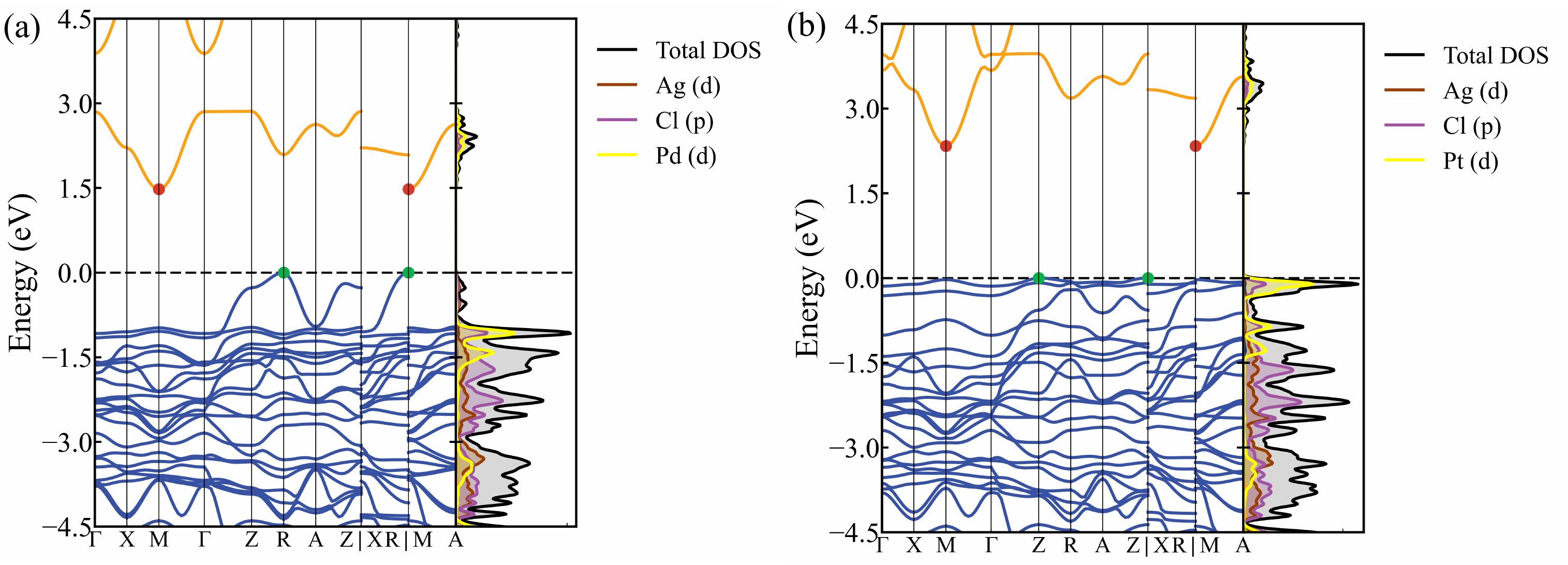}
 \caption{Electronic band structure along with the PDOS of (a) Cs$_2$AgPdCl$_5$ and (b) Cs$_2$AgPtCl$_5$ at the HSE06 level with spin orbit coupling. The band energies are scaled relative to the Fermi level.}
\label{ebs}
\end{figure}

Figures \ref{ebs}(a) and (b) present the electronic band structures along with the partial density of states (PDOS) for Cs$_2$AgPdCl$_5$ and Cs$_2$AgPtCl$_5$, respectively. We find that Cs$_2$AgPdCl$_5$ is a semiconductor with an indirect band gap of 0.51 eV, with the valence band maximum (VBM) at the R point and the conduction band minimum (CBM) at the M point of the Brillouin zone. In contrast, Cs$_2$AgPtCl$_5$ exhibits a semiconductor behavior with a direct band gap of 0.85 eV, with both the VBM and CBM located at the M point, according to the PBE approximation, Figure SI7. The higher band gap in Cs$_2$AgPtCl$_5$ compared to Cs$_2$AgPdCl$_5$ is attributed to the higher electronegativity of the Pt atom, which favors stronger ionic bonding with surrounding atoms \cite{electronegativity}. As we know that PBE's local approximation fails to predict the accurate  energy band gap, we calculated the band structure with HSE06 and including the spin orbit coupling. The electronic band gap modifies to be 1.47 eV (indirect) for Cs$_2$AgPdCl$_5$ and 2.34 eV (direct) for Cs$_2$AgPtCl$_5$ as shown in Figures \ref{ebs}(a) and (b), respectively. The VBMs at M and R point of the Brillouin zone are almost same in energy for Cs$_2$AgPtCl$_5$. We did not observe any significant change upon including spin orbit coupling as summarized in the Table \ref{bgsoc}. The small change due to spin orbit coupling can be attributed to the fact that band edges are primarily derived from hybridised states of Pd/Pt and Cl atoms with limited spin orbit splitting near the Fermi level.
 \begin{table}[h]
 \caption{HSE06 and HSE06+SOC computed band gaps for Cs$_2$AgPdCl$_5$ and Cs$_2$AgPtCl$_5$ in unit eV.}
 \centering
\begin{tabular*}{0.90\textwidth}{@{\extracolsep{\fill}}lll}
%\begin{tabular*}{0.48\textwidth}{ c c c} 
 \hline
     & HSE06 & HSE06+SOC    \\
 \hline 
  Cs$_2$AgPdCl$_5$ &  1.48    & 1.47    \\
  
   Cs$_2$AgPtCl$_5$  & 2.36  &  2.34   \\
   \hline
   \label{bgsoc}
\end{tabular*} 
\end{table} 

Table \ref{stdd} shows the effective mass of carriers along different high-symmetry directions. A comparison of their electronic band structures reveals that the conduction bands of both systems are nearly identical and highly dispersive, while the valence bands are relatively flatter. The highly dispersive nature of the conduction bands is attributed to significant contributions from the delocalized Cl-p orbitals mixed with Pd-d/Pt-d orbitals, while the flatter valence bands are primarily influenced by the localized Pd/Pt-d orbitals, as shown in the partial DOS in Figure \ref{ebs}(a) and (b). Notably, the valence band of Cs$_2$AgPtCl$_5$ is flatter compared to that of Cs$_2$AgPdCl$_5$. These extremely flat bands lead to holes with remarkably high effective mass.

In the case of Cs$_2$AgPdCl$_5$, a valley is present at the Z point in the valence band within an energy range of approximately 0.1 eV, while Cs$_2$AgPtCl$_5$ features doubly degenerate valleys at the Z point, nearly 0.03 eV below the Fermi energy. Such valleys provide additional channels for carrier conduction, significantly enhancing transport properties \cite{valley}. The partial DOS of Cs$_2$AgPdCl$_5$ shows that the majority of the contributions to the valence and conduction bands come from the Pd-d and Cl-p orbitals, with a minor contribution from the Ag-d orbitals. In contrast, the Cs atom contributes almost negligibly near the valence and conduction bands. For Cs$_2$AgPtCl$_5$, the partial DOS reveals that the valence and conduction bands are primarily formed by contributions from the Pt-d and Cl-p orbitals, with no significant contribution from the Ag-d orbitals at the valence band edge, as seen in Cs$_2$AgPdCl$_5$.

\begin{table}[h]
 \caption{Carrier effective masses along high symmetry directions M to X (M-X),M to $\Gamma$ (M-$\Gamma$) and M to A (M-A) for electrons and R to A (R-A), R to Z (R-Z) and R to X (R-X) for holes.}
 \label{stdd}
 \centering
 \begin{tabular*}{0.90\textwidth}{@{\extracolsep{\fill}}lllllll}
%\begin{tabular*}{\textwidth}{ c c c c c c c } 
 \hline
 System & \multicolumn{3}{c}{$m_e^*/m_o$} & \multicolumn{3}{c}{$m_h^*/m_o$}  \\
  \hline
              & M-X & M-$\Gamma$ & M-A &  R-A & R-Z & R-X   \\
   \hline
   Cs$_2$AgPdCl$_5$  & 0.51 & 0.47 & 0.56 &  0.3 & 1.158 & 0.6  \\
   
   Cs$_2$AgPtCl$_5$ & 0.43 & 0.41 & 0.51 & 4.61 & 5.92 & -   \\
 \hline
\end{tabular*} 
\end{table}
\subsection{Vibrational properties}
\begin{figure}[H]
 \centering
 \includegraphics[height=6cm]{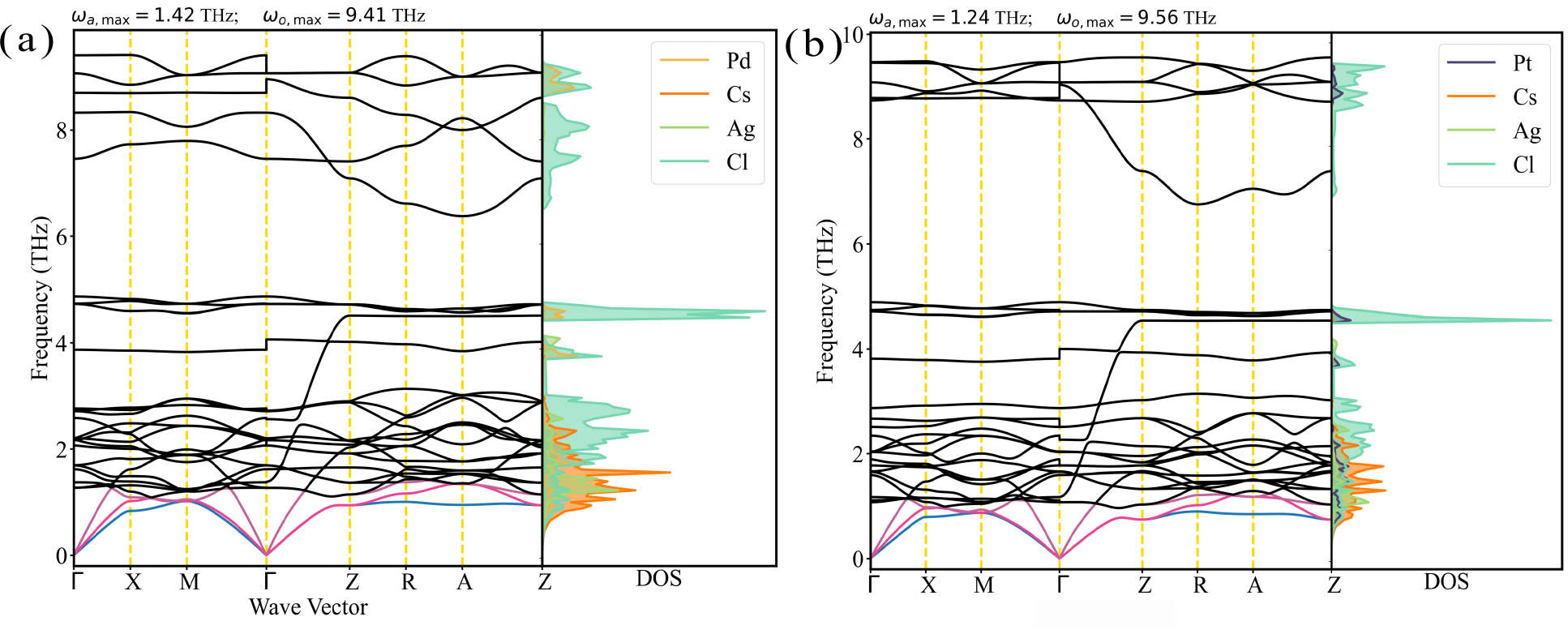}
 \caption{Phonon band structure along with partial phonon density of state (PDOS) of (a) Cs$_2$AgPdCl$_5$ and (b) Cs$_2$AgPtCl$_5$.}
\label{phbs}
\end{figure}

Figures \ref{phbs} (a) and (b) show the phonon band dispersion along the high-symmetry points of the Brillouin zones, along with the phonon partial density of states (PDOS) for Cs$_2$AgPdCl$_5$ and Cs$_2$AgPtCl$_5$, respectively. Both systems are dynamically stable, as evidenced by the absence of imaginary frequencies \cite{phase}. The discontinuity present at $\Gamma$ point is because of LO-TO splitting, Figure SI8. LO-To splitting occurs due to coulomb interaction between longitudinal and transverse phonons. Upon comparing the phonon band structures of both systems, we observe phonon softening in the acoustic phonons and subdued optical phonons in Cs$_2$AgPtCl$_5$ compared to Cs$_2$AgPdCl$_5$. However, a slight phonon hardening is observed for the high-frequency optical phonons in Cs$_2$AgPtCl$_5$. The cutoff frequencies of the acoustic modes for Cs$_2$AgPdCl$_5$ and Cs$_2$AgPtCl$_5$ are 1.42 THz and 1.24 THz, corresponding to Debye temperatures ($\Theta_D$) of 68.5 K and 60.6 K, respectively. Such low values of $\Theta_D$ are desirable for reducing heat conduction through the lattice, as they promote enhanced phonon-phonon scattering. Since acoustic phonons primarily contribute to heat conduction in the lattice, and optical phonons can indirectly influence heat conduction through electron-phonon interactions, an ultralow lattice thermal conductivity is expected for these materials. This is due to the significant overlap between acoustic and optical phonons, which further promotes phonon-phonon scattering. The projected phonon density of states (DOS) reveals that Cs makes a dominant contribution at low frequencies (up to 1.8 THz), followed by contributions from Pd, Ag, and Cl in the mid-frequency range (up to 4 THz). The light atom Cl predominantly contributes at high frequencies. The maximum optical frequencies for Cs$_2$AgPdCl$_5$ and Cs$_2$AgPtCl$_5$ are 9.41 THz and 9.56 THz, respectively. The higher optical frequency in Cs$_2$AgPtCl$_5$ compared to Cs$_2$AgPdCl$_5$ suggests enhanced phonon-phonon scattering, which contributes to the suppression of heat conduction through the lattice. Apart from this, we have performed the ab initio molecular dynamics (AIMD) simulations in the NVT ensemble using the Nose-Hoover thermostat at 800 K. As shown in Figure SI4, the crystal framework remains intact over the short timescale of 2 ps. This indicates the short time dynamical stability of the lattice at high temperature.

\subsection{Mechanical properties} 
To investigate the mechanical properties of these systems, including material stability, hardness, ductility, and brittleness, we have computed their elastic moduli. The elastic moduli characterize the system's response to applied stress in different directions \cite{elastic}. As discussed in Section \ref{strr}, these systems exhibit tetragonal structures and, consequently, possess only six independent elastic constants: C$_{11}$, C$_{12}$, C$_{13}$, C$_{33}$, C$_{44}$, and C$_{66}$. Our theoretical predictions for these elastic constants await experimental validation. The elastic constants are given in the following Table \ref{elss}.
  
 \begin{table*}
\centering
 \caption{Evaluated Elastic constants, Bulk Modulus (B), Young's modulus (E), Shear modulus (G), and Poisson ratio $\nu$ for Cs$_2$AgPdCl$_5$ and Cs$_2$AgPtCl$_5$. All the elastic constants and moduli are expressed in GPa.}
\begin{tabular*}{\textwidth}{@{\extracolsep{\fill}}lllllllllll}
%\begin{tabular*}{\textwidth}{ c c c c c c c c c c c } 
 \hline
Systems & $C_{11}$ & $C_{12}$  &$C_{13}$ & $C_{33}$ & $C_{44}$ & $C_{66}$ & B & E &  G & $\nu$  \\
 \hline 
  Cs$_2$AgPdCl$_5$ & 38.22 & 19.15 & 19.09 & 47.648 & 8.75  & 13.66 & 26.40 & 27.77 & 10.48 & 0.33 \\
  
   Cs$_2$AgPtCl$_5$  & 34.38 & 18.07 & 18.15 & 46.79 & 7.72 & 13.78 & 24.68 & 25.52 & 9.61 & 0.33 \\
   \hline
   \label{elss}
\end{tabular*} 
\end{table*} 
  From Table \ref{elss}, we observe that the Cauchy pressure ($C_p = C_{12} - C_{44}$) is positive for both systems. However, it is higher for Cs$_2$AgPdCl$_5$ than for Cs$_2$AgPtCl$_5$, indicating that Cs$_2$AgPdCl$_5$ exhibits a greater degree of ionic character \cite{cauchy_pressure}. The shear modulus (G) quantifies a material's resistance to plastic deformation, whereas the bulk modulus (B) represents its ability to resist fracture. A B/G ratio greater than 1.75 generally suggests a ductile nature, while a lower value indicates brittleness \cite{brittle}. Our calculations reveal that both systems exhibit a predominantly brittle nature.

The sound velocities of Cs$_2$AgPdCl$_5$ and Cs$_2$AgPtCl$_5$ can be derived from the computed elastic constants \cite{sound}, with the calculated values presented in Table \ref{svv}. The sound velocities of both systems are comparable, except that Cs$_2$AgPtCl$_5$ exhibits a lower sound velocity than Cs$_2$AgPdCl$_5$. This difference can be attributed to the higher atomic mass of the Pt atom. Since the Debye temperature is directly proportional to the average sound velocity, the lower sound velocity of Cs$_2$AgPtCl$_5$ results in a lower Debye temperature. \\
 
 \begin{table}[h]
 \caption{Calculated longitudinal velocity $\textrm v_l$ (m/s), transverse velocity $\textrm v_s$ (m/s), average sound velocity $\textrm v_{av}$ (m/s) and Debye temperature $\theta_D$ (K) for Cs$_2$AgPdCl$_5$ and Cs$_2$AgPtCl$_5$.}
 \centering
\begin{tabular*}{\textwidth}{@{\extracolsep{\fill}}lllll}
 \hline
Systems & v$_l$ & v$_s$  &  v$_{av}$ & $\theta_D$   \\
 \hline 
  Cs$_2$AgPdCl$_5$ & 3136.72 & 1598.23 & 1790.88 & 172.5 \\
  
   Cs$_2$AgPtCl$_5$  & 2855.93 & 1445.96 & 1620.89 & 155.4 \\
   \hline
   \label{svv}
\end{tabular*} 
\end{table}   
\subsection{Transport properties}
To assess the thermoelectric performance of these materials, we thoroughly examined their transport properties in a-b plane (in-plane) and along c direction (out of plane), such as electrical conductivity, thermal conductivity, Seebeck coefficient, power factor, and figure of merit using the PBE+SOC computed electronic band structure. We employed the HSE06+SOC calculated band gaps of 1.47 eV for Cs$_2$AgPdCl$_5$ and 2.34 eV for Cs$_2$AgPtCl$_5$. Since transport properties along in-plane direction are superior to those along c direction, The in-plane transport properties are discussed in detail below. The transport properties along c-direction are included in the SI (Figure SI10 for Cs$_2$AgPdCl$_5$, Figure SI11 for Cs$_2$AgPtCl$_5$ and Figure SI12 for zT of both system).

\begin{figure}[H]
 \centering
  \includegraphics[height=10cm]{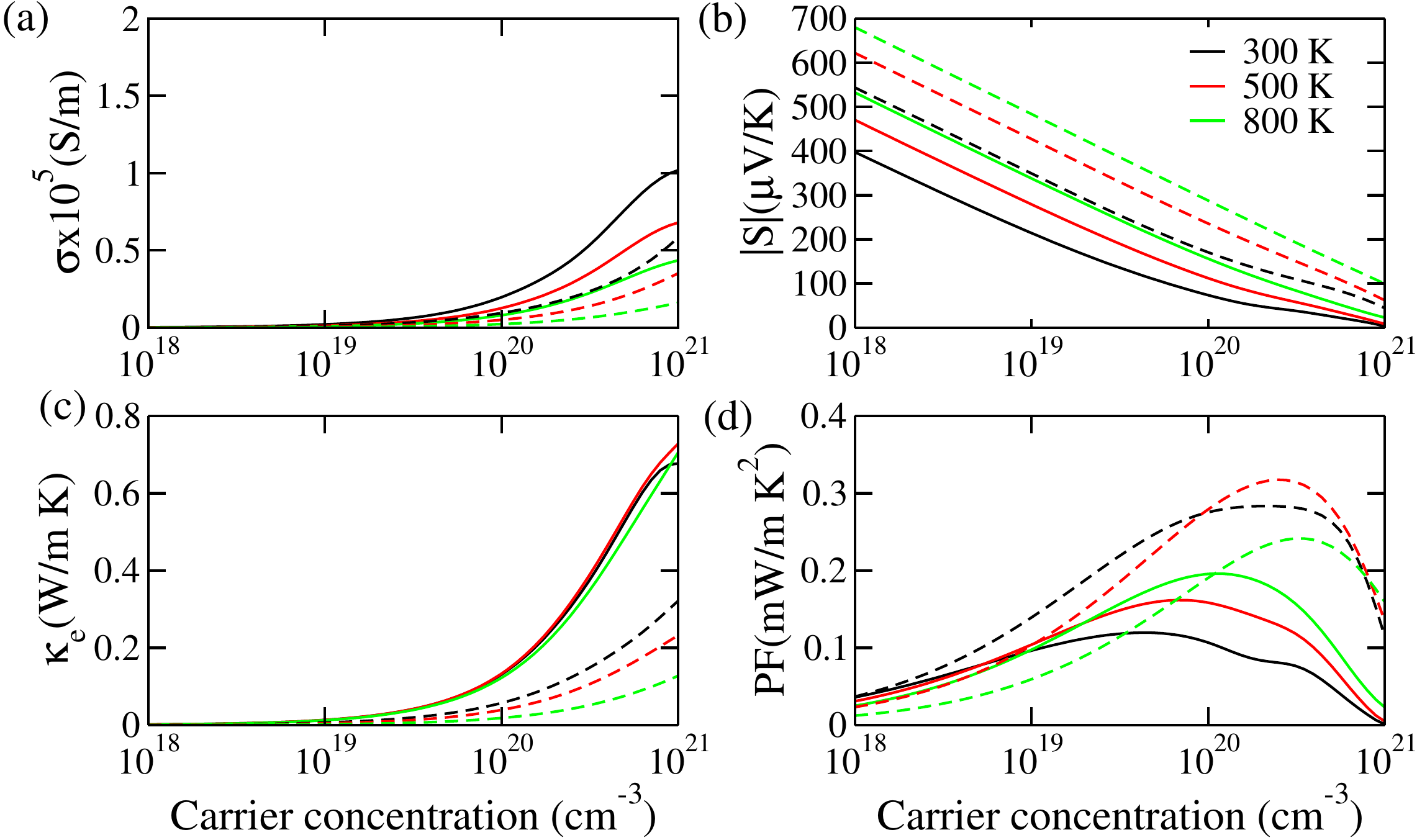}
 \caption{In plane transport properties of Cs$_2$AgPdCl$_5$ (a) electrical conductivity (b) Seebeck coefficient (c) electronic thermal conductivity (d) thermoelectric power factor}
\label{cmb_tp_pd}
\end{figure}

\begin{figure}[H]
 \centering
   \includegraphics[height=10cm]{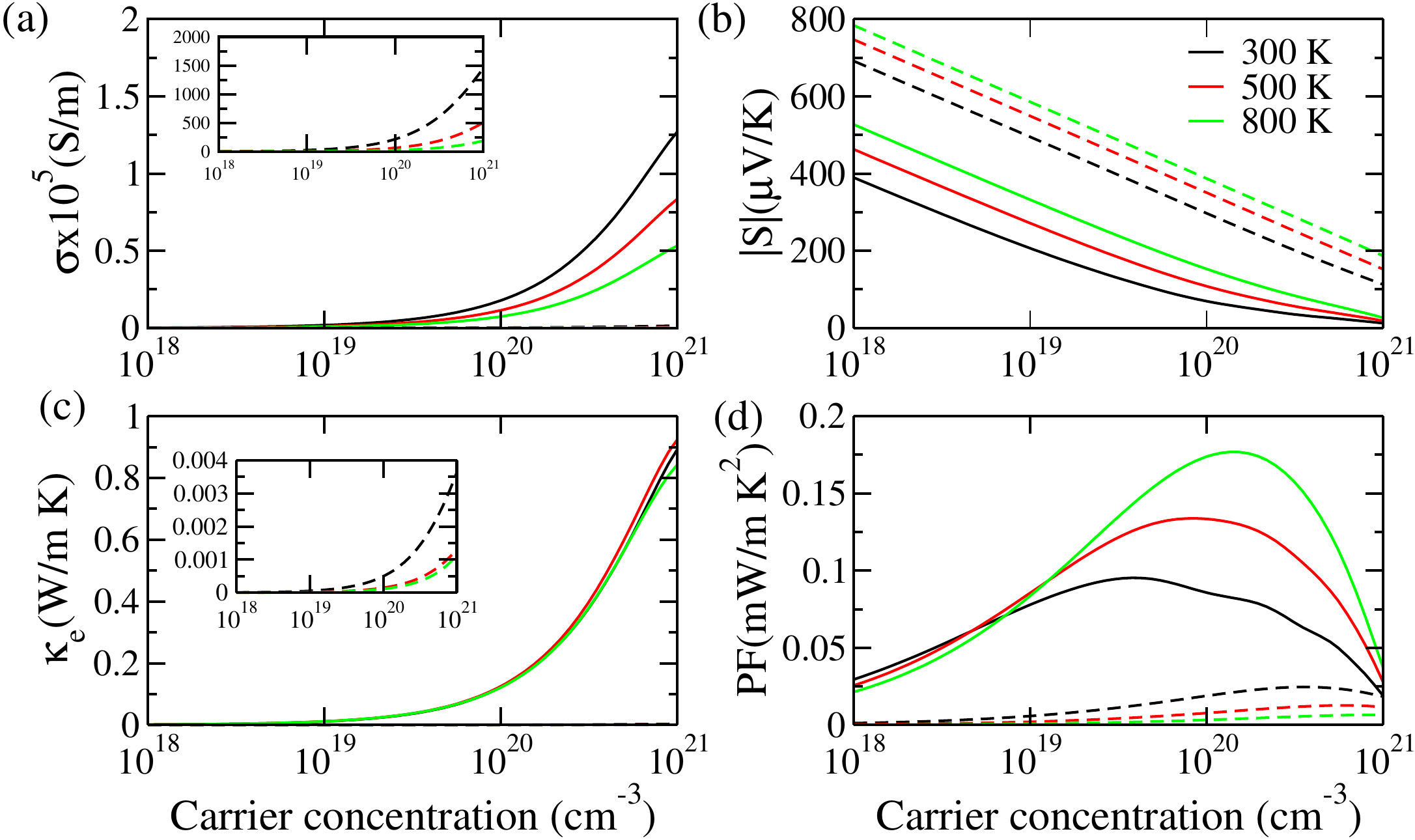}
 \caption{In plane transport properties of Cs$_2$AgPtCl$_5$ (a) electrical conductivity (b) Seebeck coefficient (c) electronic thermal conductivity (d) thermoelectric power factor}
\label{cmb_tp_pt}
\end{figure}

\begin{figure}[H]
 \centering
   \includegraphics[height=10cm]{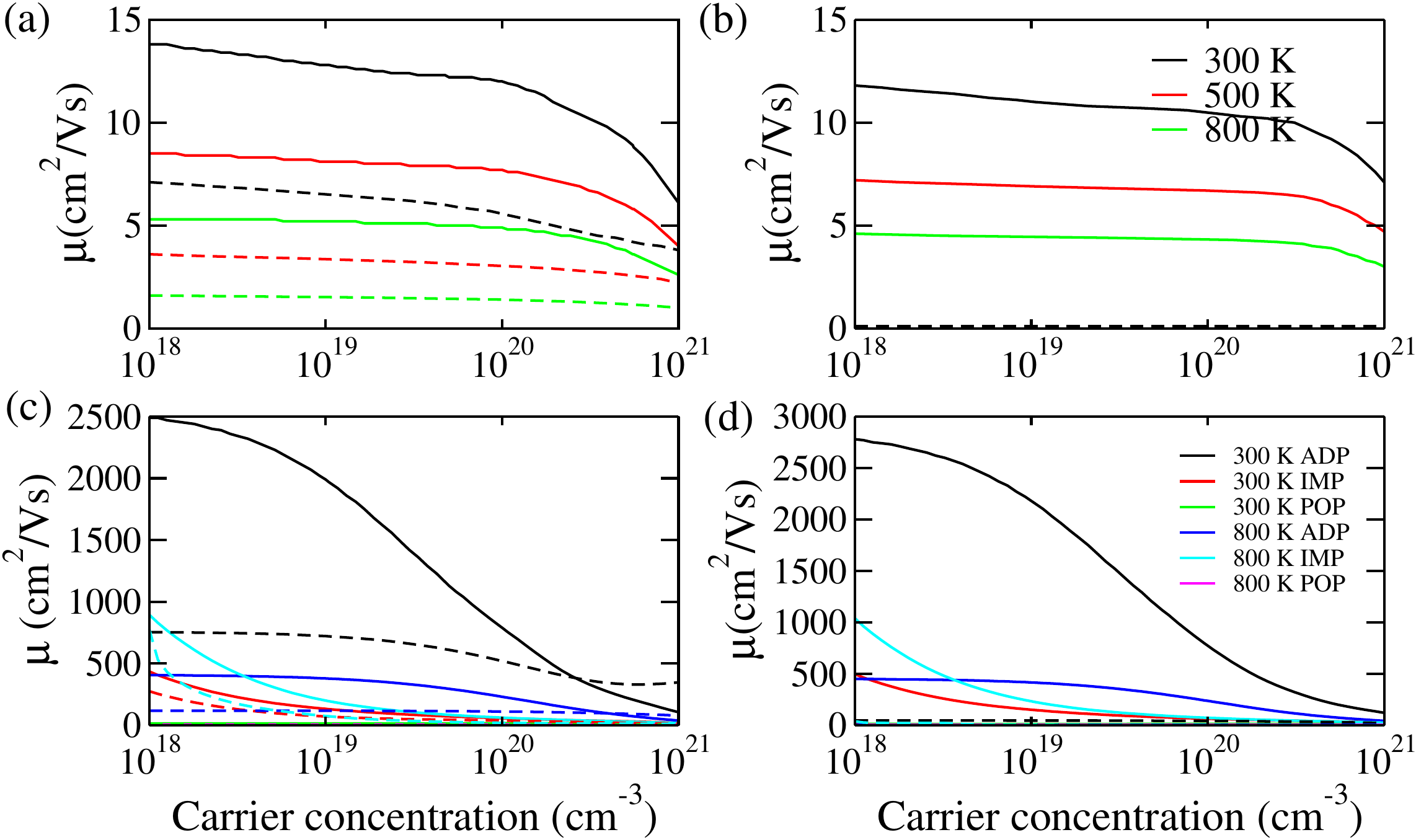}
 \caption{(a) electron and hole mobility of Cs$_2$AgPdCl$_5$, (b) electron and hole mobility of Cs$_2$AgPtCl$_5$, (c) contribution of different scattering processes in the overall mobility of Cs$_2$AgPdCl$_5$, and (d) contribution of different scattering mechanism in the overall mobility of Cs$_2$AgPtCl$_5$ for a range of temperature from 300 K to 800 K.}
\label{mob}
\end{figure}

\noindent \paragraph*{Electrical conductivity -}  The electrical conductivity ($\sigma$) of Cs$_2$AgPdCl$_5$ and Cs$_2$AgPtCl$_5$ as a function of carrier concentration over the temperatures 300 K, 500 K and 800 K is presented in Figures \ref{cmb_tp_pd}(a) and \ref{cmb_tp_pt}(a), respectively. Solid lines represent electron doping, while dashed lines indicate hole doping. In both systems, electrical conductivity is significantly higher for electron doping than for hole doping. This behavior is attributed to the presence of light carriers in highly dispersive conduction bands, in contrast to the relatively flat valence bands.
At a low electron (hole) doping concentration of $1 \times 10^{18}$ cm$^{-3}$ at 300 K, the electrical conductivity is found to be $2.27 \times 10^{2}$ S/m ($1.25 \times 10^{2}$ S/m) for Cs$_2$AgPdCl$_5$ and $1.95 \times 10^{2}$ S/m ($2.52$ S/m) for Cs$_2$AgPtCl$_5$. With increasing doping concentration, the electrical conductivity monotonically increases due to the semiconducting nature of these materials but decreases at high carrier concentrations due to an increased scattering rate. At a high electron (hole) doping concentration of $1 \times 10^{21}$ cm$^{-3}$ at 300 K, $\sigma$ is found to be $1.02 \times 10^{5}$ S/m ($5.83 \times 10^{4}$ S/m) for Cs$_2$AgPdCl$_5$ and $1.27 \times 10^{5}$ S/m ($1.45 \times 10^{3}$ S/m) for Cs$_2$AgPtCl$_5$.
A comparison of the electrical conductivity of both systems reveals that Cs$_2$AgPtCl$_5$ exhibits higher conductivity under moderate electron doping and beyond, while it is lower than that of Cs$_2$AgPdCl$_5$ for hole doping. This difference arises from the presence of lighter electrons and heavier holes in Cs$_2$AgPtCl$_5$ compared to Cs$_2$AgPdCl$_5$, as shown in Table \ref{stdd}. At a fixed carrier concentration, electrical conductivity decreases with increasing temperature. For instance, at an electron (hole) doping concentration of $1 \times 10^{18}$ cm$^{-3}$, $\sigma$ decreases from $2.27 \times 10^{2}$ S/m ($1.25 \times 10^{2}$ S/m) at 300 K to $0.88 \times 10^{2}$ S/m ($2.63 \times 10^{1}$ S/m) at 800 K for Cs$_2$AgPdCl$_5$ and from $1.95 \times 10^{2}$ S/m ($2.52$ S/m) at 300 K to $0.77 \times 10^{2}$ S/m ($0.24$ S/m) at 800 K for Cs$_2$AgPtCl$_5$. This reduction is primarily attributed to a decline in carrier mobility. Carrier mobility as a function of carrier concentration over a range of temperatures is shown in Figures \ref{mob}(a) and \ref{mob}(b) for Cs$_2$AgPdCl$_5$ and Cs$_2$AgPtCl$_5$, respectively. To further investigate mobility-limiting mechanisms, we analyze the contributions of polar optical phonon (POP) scattering, impurity scattering (IMP), and acoustic deformation potential (ADP) scattering, as depicted in Figures \ref{mob}(c) and \ref{mob}(d) at room temperature and 800 K. Our analysis reveals that IMP and POP scattering significantly limit mobility compared to the ADP mechanism. As shown in Figures \ref{mob}(c) and \ref{mob}(d), when IMP and POP scattering contributions are smaller, total mobility is primarily determined by these mechanisms, as described by Matthiessen's rule \cite{matthiessen}. Near room temperature, POP scattering is the dominant contributor, whereas IMP scattering becomes more significant at higher temperatures for both electrons and holes. Notably, holes in Cs$_2$AgPtCl$_5$ exhibit significantly lower mobility than those in Cs$_2$AgPdCl$_5$, resulting in low electrical conductivity.
It is worth noting that the electrical conductivity of these systems is comparable to that of typical thermoelectric materials, including SnSe (1000 S/m at 323 K) \cite{snse}, SnS (500 S/m at 323 K) \cite{sns}, and Cu$_2$Se (1500 S/m at 323 K) \cite{cu2se}.

\noindent \paragraph*{Electronic thermal conductivity -} The electronic thermal conductivity ($\kappa_e$) of Cs$_2$AgPdCl$_5$ and Cs$_2$AgPtCl$_5$ as a function of carrier concentration and temperature is shown in Figures \ref{cmb_tp_pd}(c) and \ref{cmb_tp_pt}(c). Solid lines represent electron doping, while dashed lines indicate hole doping. Since $\sigma$ and $\kappa_e$ are related by the Wiedemann-Franz law (Equation \ref{wd}), $\kappa_e$ follows the trend of $\sigma$ in both Cs$_2$AgPdCl$_5$ and Cs$_2$AgPtCl$_5$ for both doping types. At an electron (hole) doping concentration of $1 \times 10^{18}$ cm$^{-3}$ at 300 K, $\kappa_e$ is found to be $1.38 \times 10^{-3}$ W/mK ($0.69 \times 10^{-3}$ W/mK) for Cs$_2$AgPdCl$_5$ and $1.18 \times 10^{-3}$ W/mK ($5.88 \times 10^{-6}$ W/mK) for Cs$_2$AgPtCl$_5$. As doping concentration increases to $1 \times 10^{21}$ cm$^{-3}$, $\kappa_e$ rises to $6.77 \times 10^{-1}$ W/mK ($3.21 \times 10^{-1}$ W/mK) for Cs$_2$AgPdCl$_5$ and $0.89$ W/mK ($3.63 \times 10^{-3}$ W/mK) for Cs$_2$AgPtCl$_5$. At 800 K and an electron (hole) doping concentration of $1 \times 10^{21}$ cm$^{-3}$, $\kappa_e$ becomes $7.06 \times 10^{-1}$ W/mK ($1.27 \times 10^{-1}$ W/mK) for Cs$_2$AgPdCl$_5$ and $0.84$ W/mK ($1.14 \times 10^{-3}$ W/mK) for Cs$_2$AgPtCl$_5$. Here it is noteworthy that these $\kappa_e$ values are very small and advantageous for thermoelectric purposes.

 \begin{equation}
\frac{\kappa_e}{\sigma}=LT
\label{wd}
\end{equation}
Here $\kappa_e$ is the electronic thermal conductivity, $L$ is the Lorentz number, and $T$ is temperature. \\

\noindent \paragraph*{Seebeck coefficient -} Figures \ref{cmb_tp_pd}(b) and \ref{cmb_tp_pt}(b) illustrate the Seebeck coefficient (S) of Cs$_2$AgPdCl$_5$ and Cs$_2$AgPtCl$_5$ for temperatures 300 K, 500 K and 800 K. In both systems, the Seebeck coefficient is consistently higher for hole doping than for electron doping across the entire doping concentration range, in contrast to electrical conductivity. This trend is attributed to the significantly larger effective mass of holes compared to electrons in these materials. A comparative analysis reveals that under hole doping, Cs$_2$AgPdCl$_5$ exhibits a lower Seebeck coefficient than Cs$_2$AgPtCl$_5$, whereas under electron doping, Cs$_2$AgPdCl$_5$ exhibits a higher Seebeck coefficient than Cs$_2$AgPtCl$_5$ across all considered doping concentrations. At an electron (hole) concentration of $1 \times 10^{18}$ cm$^{-3}$ and 300 K, the Seebeck coefficient is found to be $398$ ($544$) $\mu$V/K for Cs$_2$AgPdCl$_5$ and $389$ ($691$) $\mu$V/K for Cs$_2$AgPtCl$_5$. The substantial difference in the Seebeck coefficient between electron and hole doping arises from the significant disparity in carrier effective masses, as shown in Table \ref{stdd}, in accordance with the Pisarenko relation \cite{pisarenko}. The Seebeck coefficient values obtained for these materials are notably higher than those reported for other double perovskites, such as Rb$_2$AgBiX$_6$ (X= Cl, Br) \cite{rb2agbix6}, K$_2$AgAsX$_6$ (X = halogen elements) \cite{k2agasx6}, and Cs$_2$AgBiX$_6$ (X = halogen elements) \cite{cs2biagx6}, as well as the commercially available Bi$_2$Te$_3$ \cite{bi2te3-cb}. Unlike electrical conductivity, the Seebeck coefficient decreases monotonically with increasing carrier concentration for both electron and hole doping. This decline is attributed to enhanced carrier screening, which reduces the thermally induced potential gradient. At a high electron (hole) concentration of $1 \times 10^{21}$ cm$^{-3}$ and 300 K, the Seebeck coefficient decreases to $3.42$ ($44.2$) $\mu$V/K in Cs$_2$AgPdCl$_5$ and $12.17$ ($112.46$) $\mu$V/K in Cs$_2$AgPtCl$_5$. Temperature also plays a critical role in enhancing the Seebeck coefficient, as increased scattering rates at higher temperatures contribute to its rise. At an electron (hole) concentration of $1 \times 10^{21}$ cm$^{-3}$ and 800 K, the Seebeck coefficient increases to $22.89$ ($99.45$) $\mu$V/K in Cs$_2$AgPdCl$_5$ and $26.32$ ($186.84$) $\mu$V/K in Cs$_2$AgPtCl$_5$. These high Seebeck coefficient values indicate the strong potential of these materials for thermoelectric applications.

\noindent \paragraph*{Thermoelectric power factor -} To quantify the power generation capabilities of these systems, we evaluated the thermoelectric power factor ($S^2\sigma$), which represents the combined effect of the Seebeck coefficient and electrical conductivity. The thermoelectric power factor of Cs$_2$AgPdCl$_5$ and Cs$_2$AgPtCl$_5$ as a function of carrier concentration across a range of temperatures is shown in Figures \ref{cmb_tp_pd}(d) and \ref{cmb_tp_pt}(d), respectively. At 300 K, the power factor is significantly higher for hole doping compared to electron doping. However, at 800 K, electron doping surpasses hole doping in Cs$_2$AgPdCl$_5$ up to a concentration of approximately $1.0 \times 10^{20}$ cm$^{-3}$. In contrast, for Cs$_2$AgPtCl$_5$, electron doping exhibits a higher power factor than hole doping across the entire temperature range, up to a concentration of around $1.0 \times 10^{21}$ cm$^{-3}$. This behavior is attributed to the significantly higher electrical conductivity associated with electron doping. Due to the tradeoff between electrical conductivity and the Seebeck coefficient, the thermoelectric power factor reaches a maximum at an optimized carrier concentration before declining with further doping. The optimal power factor is found to be $1.20 \times 10^{-1}$ mW/mK$^2$ at an electron concentration of $4.24 \times 10^{19}$ cm$^{-3}$ ($2.83 \times 10^{-1}$ mW/mK$^2$ at a hole concentration of $2.18 \times 10^{20}$ cm$^{-3}$) at 300 K in Cs$_2$AgPdCl$_5$. For Cs$_2$AgPtCl$_5$, the maximum power factor at 300 K is $9.53 \times 10^{-2}$ mW/mK$^2$ at an electron concentration of $3.76 \times 10^{19}$ cm$^{-3}$ ($2.46 \times 10^{-2}$ mW/mK$^2$ at a hole concentration of $3.48 \times 10^{20}$ cm$^{-3}$). With increasing temperature, these values further improve. In Cs$_2$AgPdCl$_5$, the power factor rises to $1.96 \times 10^{-1}$ mW/mK$^2$ at an electron concentration of approximately $1.08 \times 10^{20}$ cm$^{-3}$ at 800 K, while the maximum power factor for hole doping is $3.17 \times 10^{-1}$ mW/mK$^2$ at an optimized concentration of $2.45 \times 10^{20}$ cm$^{-3}$ at 500 K. In Cs$_2$AgPtCl$_5$, the maximum power factor increases to $1.77 \times 10^{-1}$ mW/mK$^2$ at an optimal electron concentration of $1.37 \times 10^{20}$ cm$^{-3}$ at 800 K. Notably, increasing the temperature does not enhance the thermoelectric power factor for hole doping in Cs$_2$AgPtCl$_5$.
Our calculations indicate that hole doping is more favourable for achieving a higher thermoelectric power factor in Cs$_2$AgPdCl$_5$, whereas electron doping is preferable in Cs$_2$AgPtCl$_5$. The obtained power factor values are comparable to those of commercially deployed thermoelectric materials, such as Bi$_2$Te$_3$ (n-type: $4.5$ mW/mK$^2$, p-type: $3.0$ mW/mK$^2$)\cite{pfbi2}, and PbTe(2.5 mW/mK$^2$) \cite{pfpbte_data}. Additionally, they are on par with several other known thermoelectric materials, including SiGe(1.6 mW/mK$^2$) \cite{pfsige}, SnSe(0.8 mW/mK$^2$)\cite{pfsnse_data}, Bi$_2$Se$_3$(2.7 mW/mK$^2$) \cite{pfbi2se3_data}, Yb$_{14}$MnSb$_{11}$ (0.6 mW/mK$^2$) \cite{pfyb}, CoSb$_3$(1.6 mW/mK$^2$) \cite{pfcosb3_data}, CsCdBr$_6$ (0.98 mW/mK$^2$) \cite{pfcscdbr6_data}, CsCdCl$_6$ (0.41 mW/mK$^2$) \cite{pfcscdbr6_data} K$_2$OsBr$_6$ (0.06 mW/mK$^2$) \cite{pfk2osbr6_data}, and Li$_2$BeAl (0.03 mW/mK$^2$) \cite{pfli2beal_data}.

\begin{figure}[H]
 \centering
   \includegraphics[height=7cm]{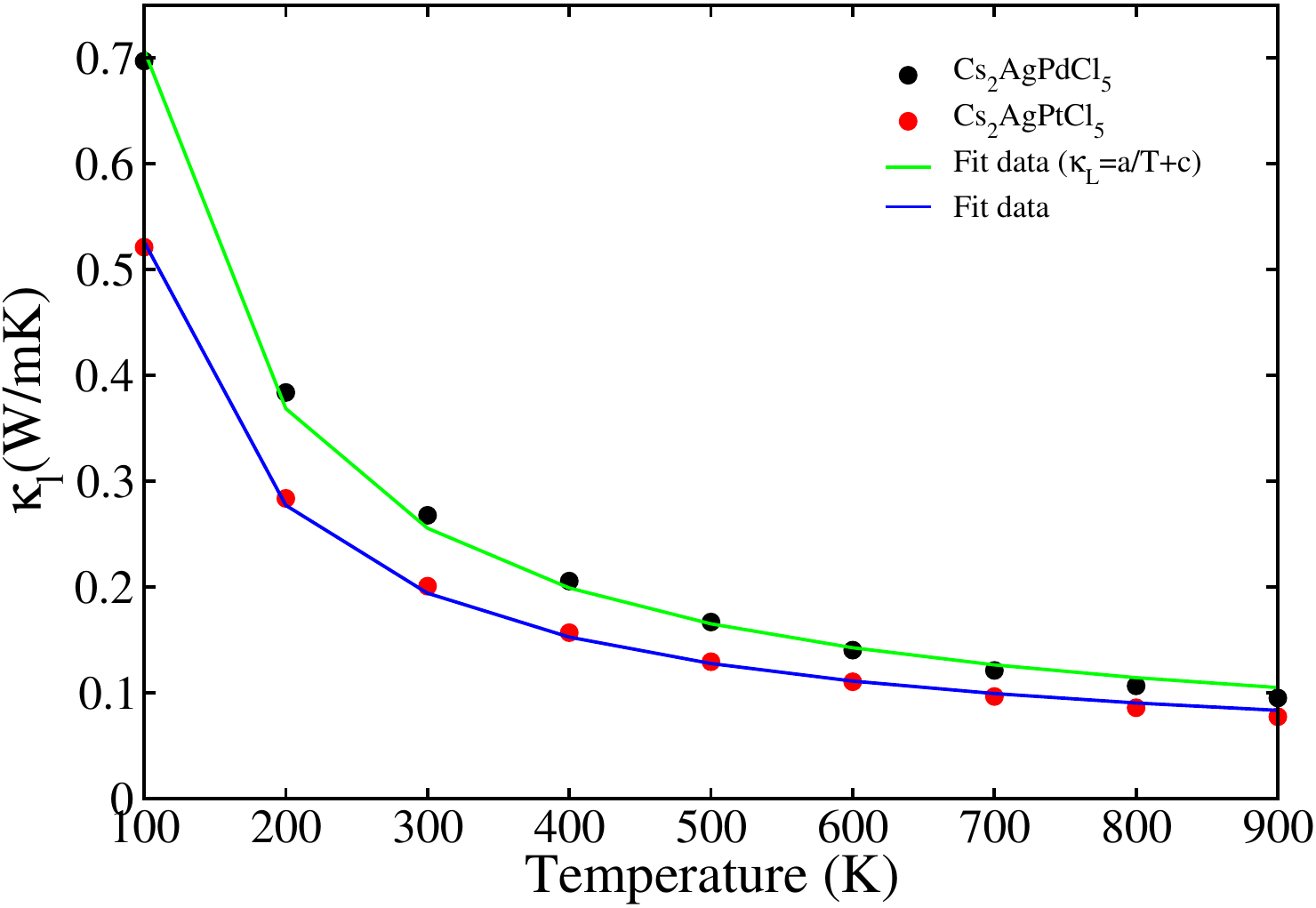}
 \caption{Lattice thermal conductivity ($\kappa_l$) as a function of the temperature of Cs$_2$AgPdCl$_5$ and Cs$_2$AgPtCl$_5$ in a-b plane}
\label{ltc}
\end{figure}

\begin{figure}[H]
 \centering
   \includegraphics[height=18cm]{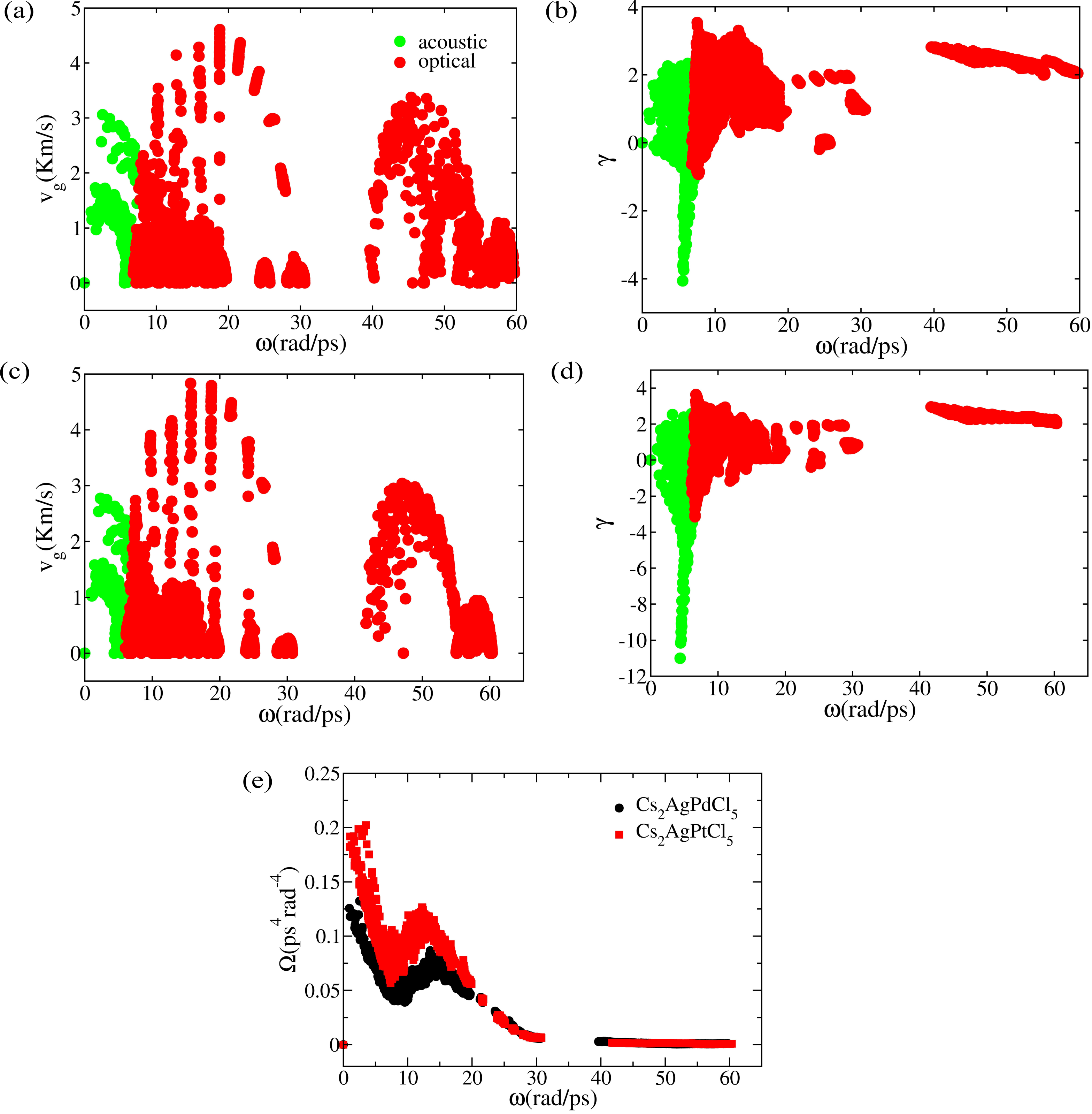}
 \caption{(a) Phonon group velocity as a function of phonon frequency and (b) gr$\ddot{\mathrm u}$neisen parameter as a function of phonon frequency at 300 K for Cs$_2$AgPdCl$_5$, (c) phonon group velocity as a function of phonon frequency and (d) gr$\ddot{\mathrm u}$neisen parameter as a function of phonon frequency at 300 K for Cs$_2$AgPtCl$_5$, (e) Three phonon scattering phase space as a function of phonon frequency at 300 K for Cs$_2$AgPdCl$_5$ and Cs$_2$AgPtCl$_5$.  }
\label{ltc1}
\end{figure}

\begin{figure}[H]
 \centering
   \includegraphics[height=8cm]{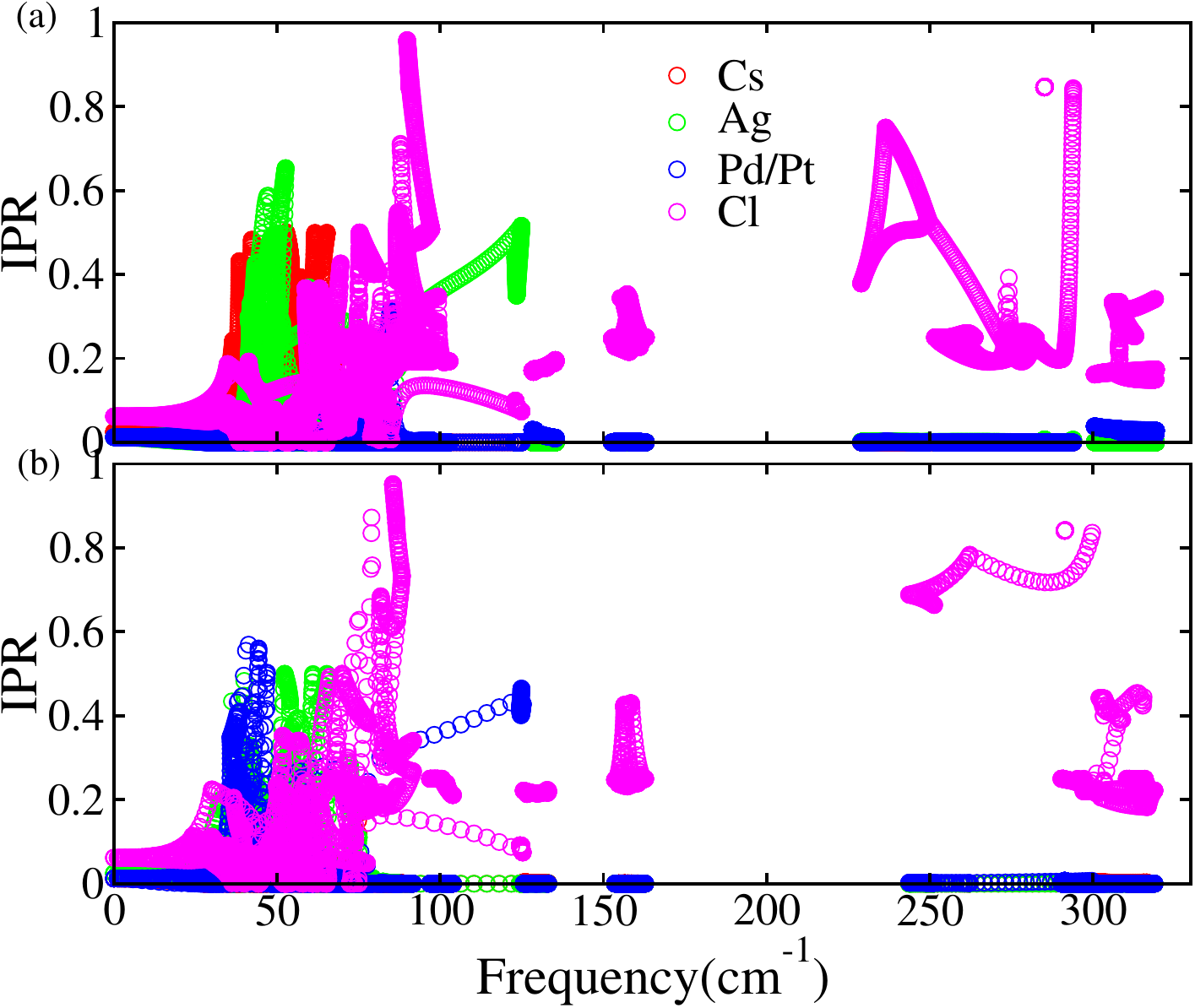}
 \caption{Projected inverse participation ratio as a function of frequency for (a) Cs$_2$AgPdCl$_5$ and (b) Cs$_2$AgPtCl$_5$}
\label{ipr}
\end{figure}

\begin{figure}[H]
 \centering
   \includegraphics[height=6cm]{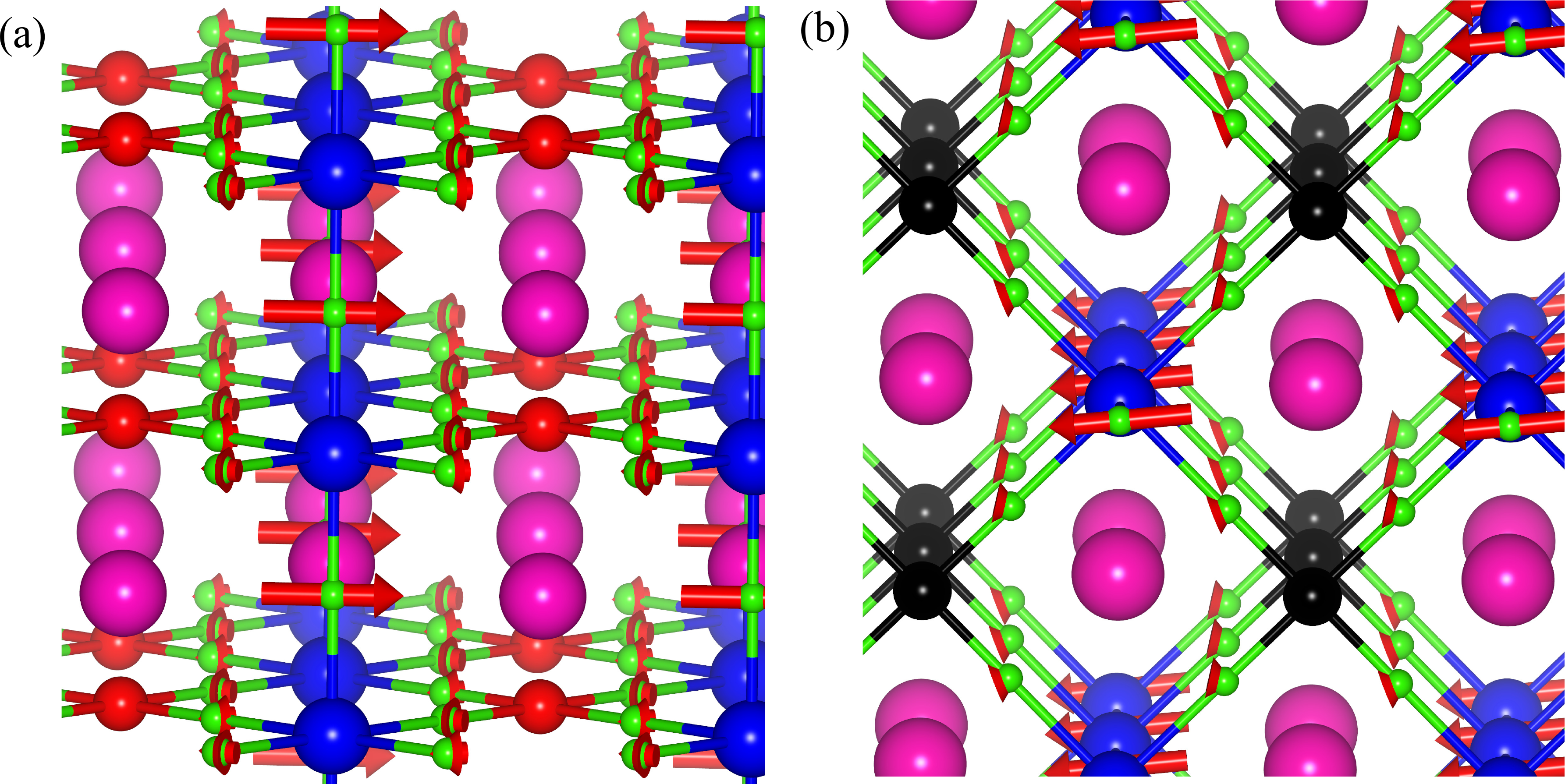}
 \caption{Displacement pattern of the mode associated with highest IPR for (a) Cs$_2$AgPdCl$_5$ and (b) Cs$_2$AgPtCl$_5$ at low optical frequency ~90 cm$^{-1}$ and in a-b plane. Here, red, black, blue, green, and magenta color shows Pd, Pt, Ag, Cl, and Cs atoms, respectively. The yellow arrows show the displacement vector for the respective atoms.}
\label{modes}
\end{figure}

\paragraph*{Lattice thermal conductivity -} Figure \ref{ltc} addresses temperature variance of lattice thermal conductivity($\kappa_l$) in a-b plane for both systems Cs$_2$AgPdCl$_5$ and Cs$_2$AgPtCl$_5$. The $\kappa_l$ along c-direction is provided in Figure SI13. The decreasing lattice thermal conductivity with increase in temperature shows its temperature dependence as T$^{-1}$ \cite{umklapp} as shown in Figure \ref{ltc}, signifying the dominance of Umklapp scattering in these systems. At room temperature, $\kappa_l$ turns out to be 0.27 (0.20) W/mk for Cs$_2$AgPdCl$_5$ (Cs$_2$AgPtCl$_5$). These ultralow values are significantly lower than those observed in some well known thermoelectric materials, such as SnSe(1.9 W/mK)\cite{klsnse}, Bi$_2$Te$_3$ (1.3 W/mK)\cite{klbi2te3}, Sb$_2$Te$_3$ (1.6 W/mK)\cite{klgete}, PbTe (~2 W/mK)\cite{pfpbte1}, GeTe (2.6 W/mK)\cite{klgete}, BaZn$_2$Sb$_2$ (1.6 W/mK) \cite{klbazn2sb2}, PbSe (1.7 W/mK) \cite{klpbse1}, SnTe (~2 W/mK) \cite{klsnte1}, and CuTaS$_3$ (~3.6 W/mK)\cite{cutas3}. These unusual ultralow $\kappa_l$ motivated us to explore the underlying cause. $\kappa_l$ mainly depends upon group velocity (v$_g$) and scattering rate ($\tau$) at room temperature, as described in equation \ref{kl}. The specific heat is constant at room temperature and above, hence it does not affect $\kappa_l$ above room temperature. The term v$_g$ depends upon the ratio of the force constant and the mass of the atom. The force constant is concerned with the bond strength in the crystal structure. Accommodating large atomic mass in a unit cell is straightforward and has been utilized to reduce $\kappa_l$ \cite{mass}. However, harnessing bond strength to reduce $\kappa_l$ is byzantine and associated with the constituent atom's electronegativity and local coordination environment. In general, the bond in higher coordination number is weaker as compared to in lower coordination number owing to longer bond length and in turn, weaker orbital overlap, manifesting Pauling's second rule \cite{pauling}. Hence, materials with the presence of octahedra and a higher coordination environment tend to have lower $\kappa_l$. For example, double perovskites Cs$_2$InAgCl$_6$ ($\kappa_l$=0.2 W m$^{-1}$K$^{-1}$ at 300 K \cite{cs2inagcl6}), Ga$_2$PdX$_6$ ($\kappa_l$=~0.1 W m$^{-1}$K$^{-1}$ at 300 K \cite{ga2pdx6}), Rb$_2$PdX$_6$ ($\kappa_l$=0.42 W m$^{-1}$K$^{-1}$ at 300 K \cite{rb2pdx6}) have ultralow $\kappa_l$. The common interesting feature among these structures is the presence of AgCl$_6$ or PdX$_6$ octahedra (X=halogen) that weakens the Ag/Pd-X bonds in these systems. In addition to the presence of such polyhedra, these systems have anti-bonding states below the Fermi energy that are expected to weaken the Ag-Cl and Pd-Cl bond further as shown in Figure \ref{cohp} \cite{weaken}. To quantify the weakness of these bonds, we evaluated the 2$^{\mathrm {nd}}$ order average force constant of Pd-Cl (Pt-Cl) to be 8.53 (10.59) eV/\AA$^2$ in Cs$_2$AgPdCl$_5$ (Cs$_2$AgPtCl$_5$), while for Ag-Cl bond, the 2$^{\mathrm {nd}}$ order force constant was obtained to be 3.57,0.25 (3.78,0.08) eV/\AA$^2$ along directions (a axis, c-axis) in Cs$_2$AgPdCl$_5$ (Cs$_2$AgPtCl$_5$). These values are significantly smaller and cause these systems to have lower values of sound velocity as given in the Table \ref{svv}. Further, Table \ref{deviation} consists of the perovskite materials with low $\kappa_l$ along with their bond length and present tilting or deviation in the system. Considering the fact that (I) The average mass of the unit cell is high (II) Both the systems Cs$_2$AgPdCl$_5$ and Cs$_2$AgPtCl$_5$ have accommodated distorted AgCl$_6$ octahedra and square planar PdCl$_4$ (PtCl$_4$) in the same unit cell, (III) The 2$^{\mathrm {nd}}$ order force constant is remarkably lower, the ultralow $\kappa_l$ at room temperature in double columnar perovskite is comprehensible. Further, to enable a direct comparison and deeper understanding of the microscopic mechanism behind the $\kappa_l$ in these systems, we examine the angular frequency dependence of group velocity, gr$\ddot{\mathrm u}$neisen parameter, and three phonon scattering phase space ($\Omega$) along with the contribution from acoustic and optical phonons for Cs$_2$AgPdCl$_5$ and Cs$_2$AgPtCl$_5$. The higher group velocity of acoustic phonons in Cs$_2$AgPdCl$_5$ over Cs$_2$AgPtCl$_5$ corroborates slightly higher $\kappa_l$ in Cs$_2$AgPdCl$_5$ (Figure \ref{ltc1}(a) and (c)). The higher group velocity of low lying optical phonons indicates their strong interaction with acoustic phonons in both systems. The gr$\ddot{\mathrm u}$neisen parameter variance with angular frequency is shown in Figure \ref{ltc1}(b) and (d) for Cs$_2$AgPdCl$_5$ and Cs$_2$AgPtCl$_5$, respectively. Gr$\ddot{\mathrm u}$neisen parameter indicates the degree of anharmonicity and negatively affects $\kappa_l$ in the system \cite{anharmonic}. The higher gr$\ddot{\mathrm u}$neisen parameter of acoustic phonons in Cs$_2$AgPtCl$_5$ further helps in subsiding its $\kappa_l$ as compared to Cs$_2$AgPdCl$_5$. This slightly higher Gr$\ddot{\mathrm u}$neisen parameter in Cs$_2$AgPtCl$_5$ stems from the distortion present in AgCl$_6$ octahedra (see Table\ref{tabb1}).
The macroscopic average Gr$\ddot{\mathrm u}$neisen parameter as function of temperature is provided in Figure SI9. Figure \ref{ltc1}(e) shows the angular frequency dependence of $\Omega$ in Cs$_2$AgPdCl$_5$ and Cs$_2$AgPtCl$_5$ at 300 K, respectively. Large value of $\Omega$ indicates more number of available states for phonon-phonon scattering. $\Omega$ shows non monotonic behaviour with increase in angular frequency. Comparatively, presence of larger $\Omega$ in Cs$_2$AgPtCl$_5$ over Cs$_2$AgPdCl$_5$ in the mid frequency range maximizes the phonon-phonon scattering that further upholds lower $\kappa_l$ in Cs$_2$AgPtCl$_5$. 
In addition to this, we analyzed the inverse participation ratio (IPR) to substantiate our claim of suppression of $\kappa_l$ caused by distorted polyhedra. The IPR is defined in equation \ref{ipreq}, where N denotes the total number of atoms and $u_{i\alpha k}$ indicates normalized eigenvector for $i^{th}$ atom along $\alpha$ direction and for $k^{th}$ phonon mode.  IPR help us identify the dominant localized modes that significantly hinder the phonon conduction. In essence, the IPR approaches to zero perfectly delocalized mode is zero and for a perfectly localized mode is one. Figure \ref{ipr} (a) and (b) illustrate the projected IPR as function of frequency for Cs$_2$AgPdCl$_5$ and Cs$_2$AgPtCl$_5$, respectively. The highest IPR value in both the systems is attributed to the Chlorine (Cl) atoms of the distorted Ag-Cl octahedra in the low lying optical frequency zone near 89 cm$^{-1}$ for Cs$_2$AgPdCl$_5$ and 82 cm$^{-1}$ for Cs$_2$AgPtCl$_5$. Figure \ref{modes} shows the displacement pattern of the phonon mode associated with these high IPR. The asymmetrical vibrational pattern of the stretched Cl atoms as compared to compressed Cl atoms are inducing the enhanced phonon scattering. This observation confirms the crucial role of distorted octahedra in impeding the phonon conduction. With increase in temperature, the $\kappa_l$'s ultralow values reduce further to be 0.10 (0.09) in Cs$_2$AgPdCl$_5$(Cs$_2$AgPtCl$_5$) at 800 K owing to increased phonon-phonon interaction at high temperature. Table:\ref{klt} shows the $\kappa_l$ of other double perovskite materials. These small values of $\kappa_l$ are much desirable for thermoelectric applications. 
%The reported values of $\kappa_l$, obtained using third-order anharmonicity, change when computed using fourth-order force constants as given in Table S1 and Table S2 in SI, which could lead to small variation in the estimated zT values. However, it is not expected to alter the qualitative conclusion of this study. Here it is worth mentioning that the values in Table S1 and Table S2 are computed using $3\times 3\times 4$ grid are not converged due to lack of computational resources.

 \begin{table}[h]
 \caption{Lattice thermal conductivity of other known perovskite materials at 300 K}
 \centering
\begin{tabular*}{\textwidth}{@{\extracolsep{\fill}}ll}
 \hline
Systems & $\kappa_l$ (W/mK)   \\
 \hline 
  CsSnI$_3$ & 0.38\cite{csgei3}  \\
  
   CsGeI$_3$  & 0.58\cite{csgei3}  \\
   
   Cs$_2$BiAgCl$_6$ & 0.08\cite{cs2biagcl6} \\
   
   Cs$_2$SnI$_6$  &  ~0.14-0.3\cite{cs2sni6, cssni6}  \\
   
   Cs$_2$PtI$_6$   &   0.15\cite{cs2pti6} \\
   
   Cs$_2$AgBiBr$_6$ & 0.35 \cite{csagbibr} \\
   
   Cs$_2$NaInCl$_6$ & 0.43 \cite{csnaincl} \\
   
   CsPbI$_3$ & 0.45 \cite{cspbx} \\
   
   CsPbBr$_3$ &  0.42 \cite{cspbx} \\
   
   Cs$_2$AgPdCl$_5$ & 0.27 \\
   
   Cs$_2$AgPtCl$_5$ & 0.20 \\
   
   \hline
   \label{klt}  
\end{tabular*} 
\end{table}   

 \begin{table*}[h]
\centering
 \caption{Perovskite materials with bond length and deviation along with $\kappa_l$ at 300 K }
\begin{tabular*}{\textwidth}{@{\extracolsep{\fill}}llll}
 \hline
Systems & Bond length (\AA) & Bond tilting/deviation  & $\kappa_l$ (W/mK)   \\
 \hline 
  SrTiO$_3$ \cite{sr1, sr2} & Ti-O=1.96 &  No tilting  & 7.6  \\
  
  CsPbBr$_3$ \cite{cspbx, cspbx1} & Pb-Br= 2.99 & octahedral tilt & 0.43   \\

   Cs$_2$BiAgBr$_6$ \cite{csagbibr, csagbibr1} & Ag-Br=2.8, Bi-Br=2.9 & octahedral tilt in AgBr$_6$/BiBr$_6$ & 0.35  \\
   
   Cs$_2$PtI$_6$ \cite{cs2pti6}  &   Pt-I=2.8 & I-Pt-I <  180$\deg$ & 0.3 \\
   
   \hline
   \label{deviation} 
\end{tabular*} 
\end{table*}

\begin{equation}
\mathrm{IPR}= \sum_{i=1}^{N} \left( \sum_{\alpha=1}^{3} u_{i\alpha,k}^{2} \right)^{2}
\label{ipreq}
\end{equation}

\begin{figure}[H]
 \centering
   \includegraphics[height=10cm]{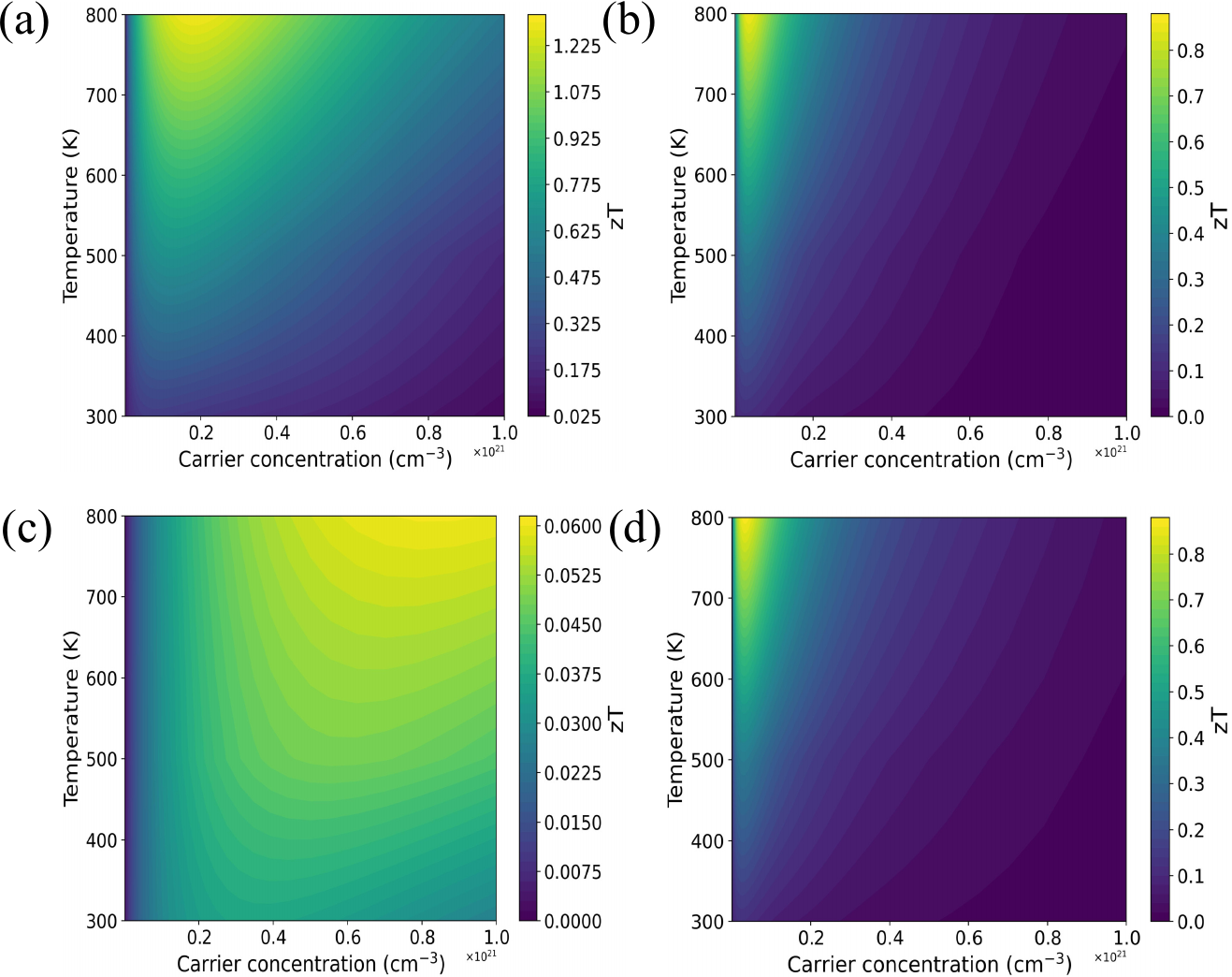}
 \caption{Figure of merit (zT) of (a) p-type (b) n-type of Cs$_2$AgPdCl$_5$ and (c) p-type (d) n-type of Cs$_2$AgPtCl$_5$  }
\label{pdzt}
\end{figure}

\paragraph*{Figure of merit } We assessed the thermoelectric figure of merit (zT) of the systems Cs$_2$AgPdCl$_5$ and Cs$_2$AgPtCl$_5$ by utilizing the above calculated electronic and transport properties. Figure \ref{pdzt} (a) and (b)[Figure \ref{pdzt}(c) and (d)] show the zT of the system Cs$_2$AgPdCl$_5$ [Cs$_2$AgPtCl$_5$] for hole doping and electron doping, respectively, as a function of carrier concentration and over temperatures 300 K, 500 K and 800 K. Our results demonstrate that both types of doping has remarkable zT in Cs$_2$AgPdCl$_5$. However in Cs$_2$AgPtCl$_5$, n-type doping is favourable over p-type doping. Interestingly, the hole doping in Cs$_2$AgPdCl$_5$ has shown favourable thermoelectric performance with zT 1.30 at 800 K with hole doping of $1.94 \times 10^{20}$ cm$^{-3}$ and 0.86 at electron doping $3.76 \times 10^{19}$ cm$^{-3}$ and 800 K. These zT values are competitive to some of the art thermoelectric materials namely SnTe (1 at 900 K) \cite{klsnte1}, PbTe (1.96 at 700 K) \cite{ztpbte}, Bi$_2$Te$_3$ (~1.03 at 400 K)\cite{ztbi2te3}, Bi$_2$Se$_3$ (1.14 at 300 K) \cite{pfbi2se3_data}, CoSb$_3$(0.43 at 600 K) \cite{pfcosb3_data}, Yb$_{14}$MnSb$_{11}$ (~1 at 1200 K)\cite{pfyb}, SiGe (1.84 at 1100 K) \cite{pfsige}, CsCdBr$_6$ (1.16 at 900 K) \cite{pfcscdbr6_data} and Cs$_2$InAgCl$_6$ (0.74 at 700 K) \cite{cs2inagcl6}. This elevated zT for hole doping originates from its high electrical conductivity and Seebeck coefficient. In essence, it is the result of its suitable valence band curvature along with the presence of band degeneracy. Though the hole doping in Cs$_2$AgPtCl$_5$ has not shown significant improvement in zT, the electron doping has shown substantial improvement in zT to be 0.87 at optimum concentration $3.52 \times 10^{19}$ cm$^{-3}$ and 800 K. Hence, these materials show excellent thermoelectric performance for both types of doping. Here we emphasize that while highest predicted thermoelectric performance occurs at 800 K for hole doping in Cs$_2$AgPdCl$_5$,  experimental studies report the onset of mass loss at this elevated temperatures, particularly associated with halogen vacancy formation. Hence, these results should be regarded as an upper limit estimate and it may require controlled atmosphere or additional stabilization strategies.

 \begin{table*}[h]
 \centering
\caption{Key thermoelectric parameters- band gap($E_g$), lattice thermal conductivity($\kappa_l$), peak zT, optimum carrier concentration($n_{op}$).}
\begin{tabular*}{\textwidth}{@{\extracolsep{\fill}}lllll}
 \hline
Systems & $E_g$(eV) & $\kappa_l$(W/mK)  &  peak zT & $n_{op}(cm^{-3})$   \\
 \hline 
 PbTe \cite{pbte111}  &    0.28-0.31   & $\sim$ 2   &  1.0   &  $\sim$ 2 $\times$ 10$^{19}$(holes)      \\

 Bi$_2$Te$_3$ \cite{bi2te3111}  &   0.16  &  $\sim$ 1.2   &  $\sim$  1.0   &  $\sim$ 10$^{19}$(holes)  \\
 
 Yb$_{14}$MnSb$_{11}$ \cite{ybmnsb111}  &  $\sim$ 0.4    &   $\sim$ 0.4    & $\sim$ 1.3        &   $\sim$ 10$^{19}$ (holes)               \\
 
 SnSe \cite{snse111}   &  $\sim$ 0.9     &  $\sim$ 0.23    &  $\sim$ 2.6      &   $\sim$ 10$^{17}$     \\
 
 CoSb$_3$ \cite{cosb3111}     &  $\sim$ 0.2-0.3        &   $\sim$ 10        &  $\sim$ 0.65             &     $\sim$ 10$^{19}$ (holes)                  \\
 
 Cs$_2$InAgCl$_6$  \cite{csinagcl111}  &   $\sim$ 3.3        &  $\sim$ 0.2        &    $\sim$ 0.74    &  $\sim$ 10$^{19}$ (holes) \\
 
 CuTaS$_3$  \cite{cutas3}       &   $\sim$ 1.0         &     $\sim$ 0.6        &   $\sim$ 0.9         &    $\sim$ 10$^{19}$ (holes) \\
 
  Cs$_2$AgPdCl$_5$ (This work) & 1.47 & 0.10 & 1.30  & 1.94 $\times$ 10$^{20}$ (holes) \\
  
   Cs$_2$AgPtCl$_5$ (This work)  & 2.34 & 0.09 & 0.87 & 3.52 $\times$ 10$^{19}$(electrons) \\
   \hline
   \label{compare_table} 
\end{tabular*} 
\end{table*}   

\section{Conclusions}  
\label{conc}
   In summary, this study presents the detailed, systematic analysis of the electronic, mechanical, vibrational and thermoelectric properties of 1D columnar halide perovskites Cs$_2$AgPdCl$_5$ and Cs$_2$AgPtCl$_5$. Both systems are found to be dynamically and mechanically stable. The presence of significant distortion in [AgCl$_6$] octahedra introduces bond anharmonicity and lowers the $\kappa_l$ notably in these systems. This is also evident from the presence of the prominent broad antibonding COHP peaks of Ag-Cl and Pd-Cl/Pt-Cl bonds present in the system that weakens the bond and leads to minimal $\kappa_l$. For example, $\kappa_l$ of Cs$_2$AgPdCl$_5$ and Cs$_2$AgPtCl$_5$ was found to be 0.27 and 0.20 at 300 K. Along with the ultralow $\kappa_l$, these systems have moderate band gaps and suitable electronic band curvature combined with the presence of band valleys that improve its thermoelectric power factor and in turn, zT of these systems. In general, a zT greater than 1 is desirable for thermoelectric applications. These systems exhibit remarkably impressive thermoelectric performance with zT $\sim$ 1.0 for both types of doping eg, we obtain zT to be 1.30 for Cs$_2$AgPdCl$_5$ at an optimum hole concentration $1.94 \times 10^{20}$ cm$^{-3}$ and 0.86 at electron concentration $3.76 \times 10^{19}$ cm$^{-3}$ at 800 K, whereas for Cs$_2$AgPtCl$_5$, we find zT to be 0.87 at an optimum electron concentration of $3.52 \times 10^{19}$ cm$^{-3}$ at 800 K for electron doping. With these compelling properties, these materials are anticipated to be promising materials for applications in thermoelectric devices.

\section*{Author contributions}
Heena: Investigation, methodology, data curation, visualization, writing the original draft. Vineet Kumar Pandey: Investigation, writing the original draft, systematic analysis, validation, data curation, visualization. Ambesh dixit: systematic analysis, resources, writing and editing. Anver Aziz: Resources. Sung Gu Kang: Resources. K.C.Bhamu: conceptualization, supervision, validation, writing and editing, resources.

\section*{Conflicts of interest}
The authors claim no competing interests to report.

\section*{Data availability}
Upon valid request, the corresponding author can grant access to the data pertaining to this study
\section*{Acknowledgements}
 The authors acknowledge the National Supercomputing Mission (NSM) for providing computing resources of PARAM RUDRA at C-DAC, NEW DELHI, which is implemented by C-DAC and supported by the Ministry of Electronics and information Technology (MeiTY) and the department of Science and Technology (DST), Government of India. The authors also acknowledge the High Performance Computing (HPC) facilities at the Indian Institute of Technology Jodhpur, India, and the University of Ulsan, South Korea, for additional computational support. 

\bibliographystyle{unsrt}
\bibliography{PdX}
\newpage
\section*{Supporting information}
\textbf{AMSET Input Parameters for Cs$_2$AgPdCl$_5$ and Cs$_2$AgPtCl$_5$}

\subsection*{1. Cs$_2$AgPdCl$_5$}

\textbf{Interpolation factor:} 5

\subsubsection*{High-frequency dielectric tensor ($\varepsilon_{\infty}$)}
\[
\begin{bmatrix}
5.314976 & 0 & 0 \\
0 & 5.314976 & 0 \\
0 & 0 & 3.832
\end{bmatrix}
\]

\subsubsection*{Static dielectric tensor ($\varepsilon_{0}$)}
\[
\begin{bmatrix}
11.111574 & -0.000393 & 0.000038 \\
-0.000393 & 12.105745 & -0.000054 \\
0.000038 & -0.000054 & 8.868483
\end{bmatrix}
\]

\subsubsection*{Elastic tensor (GPa)}
\[
\begin{bmatrix}
38.221 & 19.154 & 19.089 & 0 & 0 & 0 \\
19.154 & 38.221 & 19.089 & 0 & 0 & 0 \\
19.089 & 19.089 & 47.648 & 0 & 0 & 0 \\
0 & 0 & 0 & 8.747 & 0 & 0 \\
0 & 0 & 0 & 0 & 8.747 & 0 \\
0 & 0 & 0 & 0 & 0 & 8.747
\end{bmatrix}
\]

\textbf{POP frequency:} 3.11 THz \\
\textbf{fd\_tol:} 0.005 \\
\textbf{dos\_estep:} 0.002 eV 

%------------------------------------------------
\subsection*{2. Cs$_2$AgPtCl$_5$}

\subsubsection*{High-frequency dielectric tensor ($\varepsilon_{\infty}$)}
\[
\begin{bmatrix}
5.314976 & 0 & 0 \\
0 & 5.314976 & 0 \\
0 & 0 & 3.832753
\end{bmatrix}
\]

\subsubsection*{Static dielectric tensor ($\varepsilon_{0}$)}
\[
\begin{bmatrix}
11.782089 & 0.000140 & 0.000036 \\
0.000140 & 12.089111 & -0.000005 \\
0.000036 & -0.000005 & 8.941241
\end{bmatrix}
\]

\subsubsection*{Elastic tensor (GPa)}
\[
\begin{bmatrix}
34.377 & 18.068 & 18.146 & 0 & 0 & 0 \\
18.068 & 34.377 & 18.146 & 0 & 0 & 0 \\
18.146 & 18.146 & 46.792 & 0 & 0 & 0 \\
0 & 0 & 0 & 7.721 & 0 & 0 \\
0 & 0 & 0 & 0 & 7.721 & 0 \\
0 & 0 & 0 & 0 & 0 & 13.776
\end{bmatrix}
\]

\textbf{POP frequency:} 2.92 THz \\
\textbf{fd\_tol:} 0.005 \\
\textbf{dos\_estep:} 0.002 eV

%------------------------------------------------
%\section*{Supplementary Figures}

%\begin{itemize}

%\item \textbf{Figure S1:} Transport properties of Cs$_2$AgPdCl$_5$ (a) Electrical conductivity, (b) Seebeck coefficient, (c) Electronic thermal conductivity, (d) Power factor with different interpolation factors.
\begin{figure}[H]
 \centering
   \includegraphics[height=8cm]{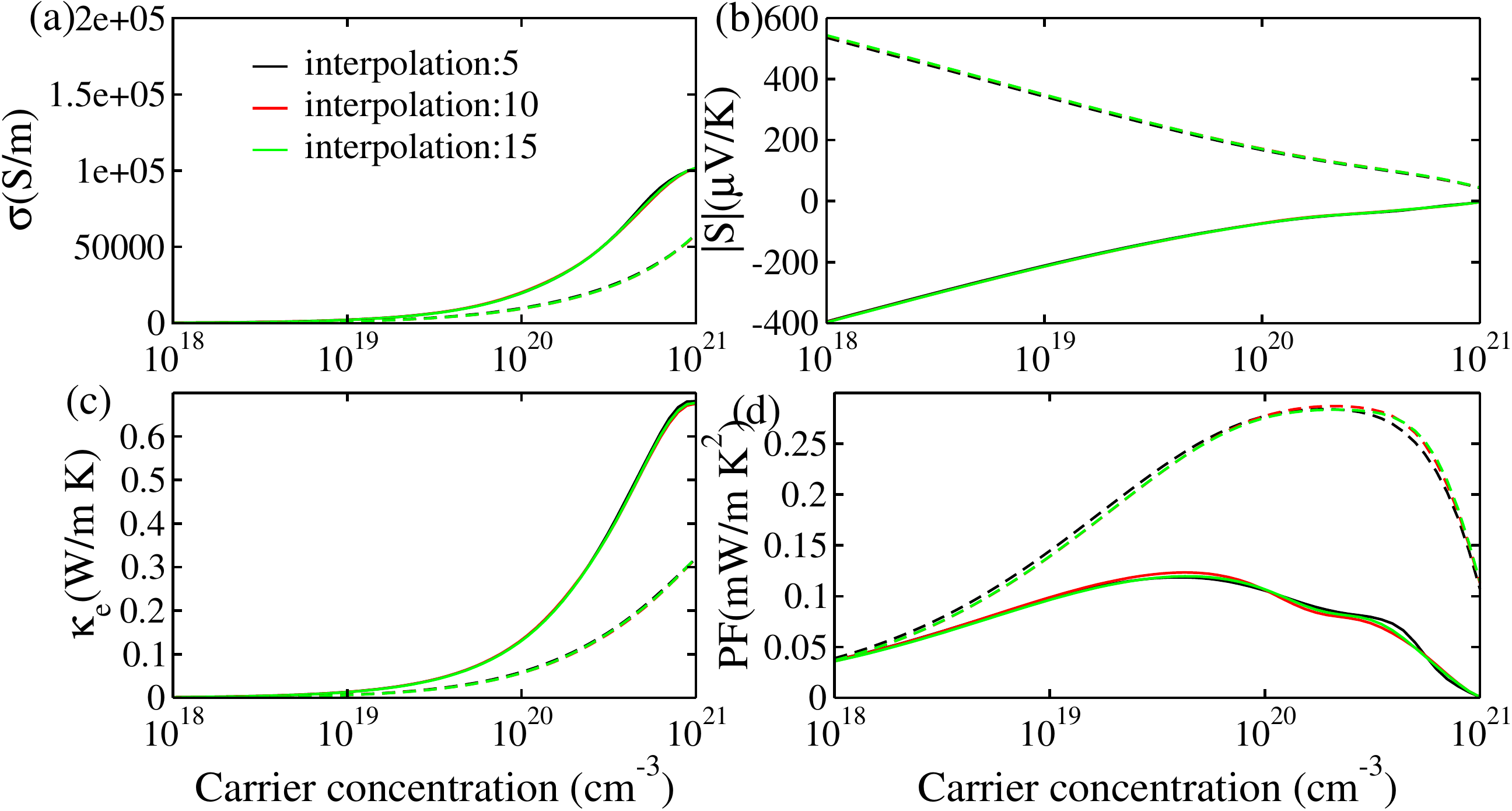}
% \caption{\mathrm{Figure S1: Transport properties of Cs$_2$AgPdCl$_5$ (a) Electrical conductivity, (b) Seebeck coefficient, (c) Electronic thermal conductivity, (d) Power factor with different interpolation factors}  }
%\label{pdzt}
\end{figure}
Figure SI1: Transport properties of Cs$_2$AgPdCl$_5$ (a) Electrical conductivity, (b) Seebeck coefficient, (c) Electronic thermal conductivity, (d) Power factor with different interpolation factors
%\item \textbf{Figure S2:} Transport properties of Cs$_2$AgPtCl$_5$ (a) Electrical conductivity, (b) Seebeck coefficient, (c) Electronic thermal conductivity, (d) Power factor with different interpolation factors.
\begin{figure}[H]
 \centering
   \includegraphics[height=8cm]{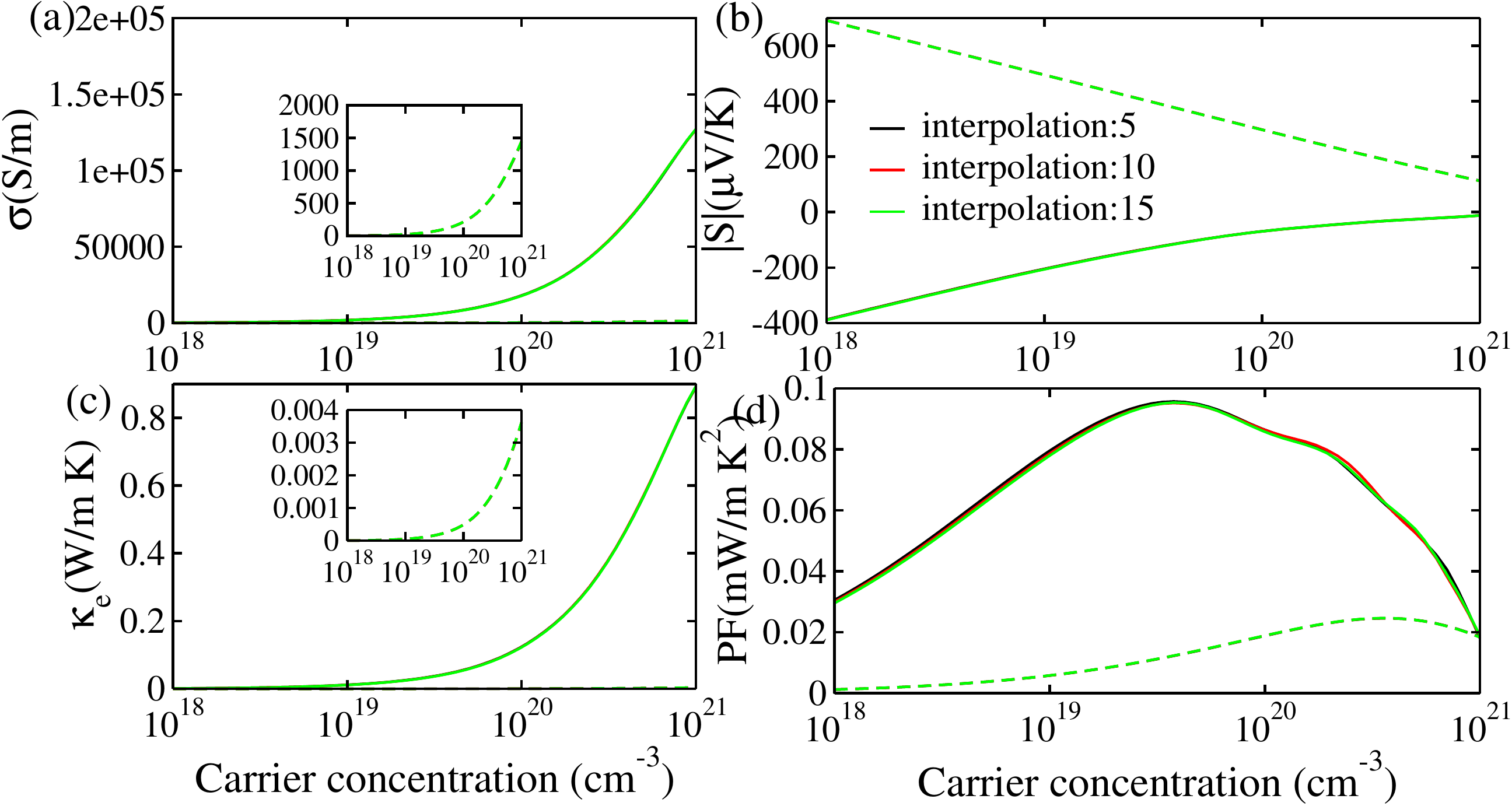}
% \caption{\mathrm{Figure of merit (zT) of (a) p-type (b) n-type of Cs$_2$AgPdCl$_5$ and (c) p-type (d) n-type of Cs$_2$AgPtCl$_5$}  }
%\label{pdzt}
\end{figure}
Figure SI2: Transport properties of Cs$_2$AgPtCl$_5$ (a) Electrical conductivity, (b) Seebeck coefficient, (c) Electronic thermal conductivity, (d) Power factor with different interpolation factors.
%\item \textbf{Figure S3:} Convergence of lattice thermal conductivity with respect to q-grid ($n_x \times n_y \times %n_z$).
\begin{figure}[H]
 \centering
   \includegraphics[height=7cm]{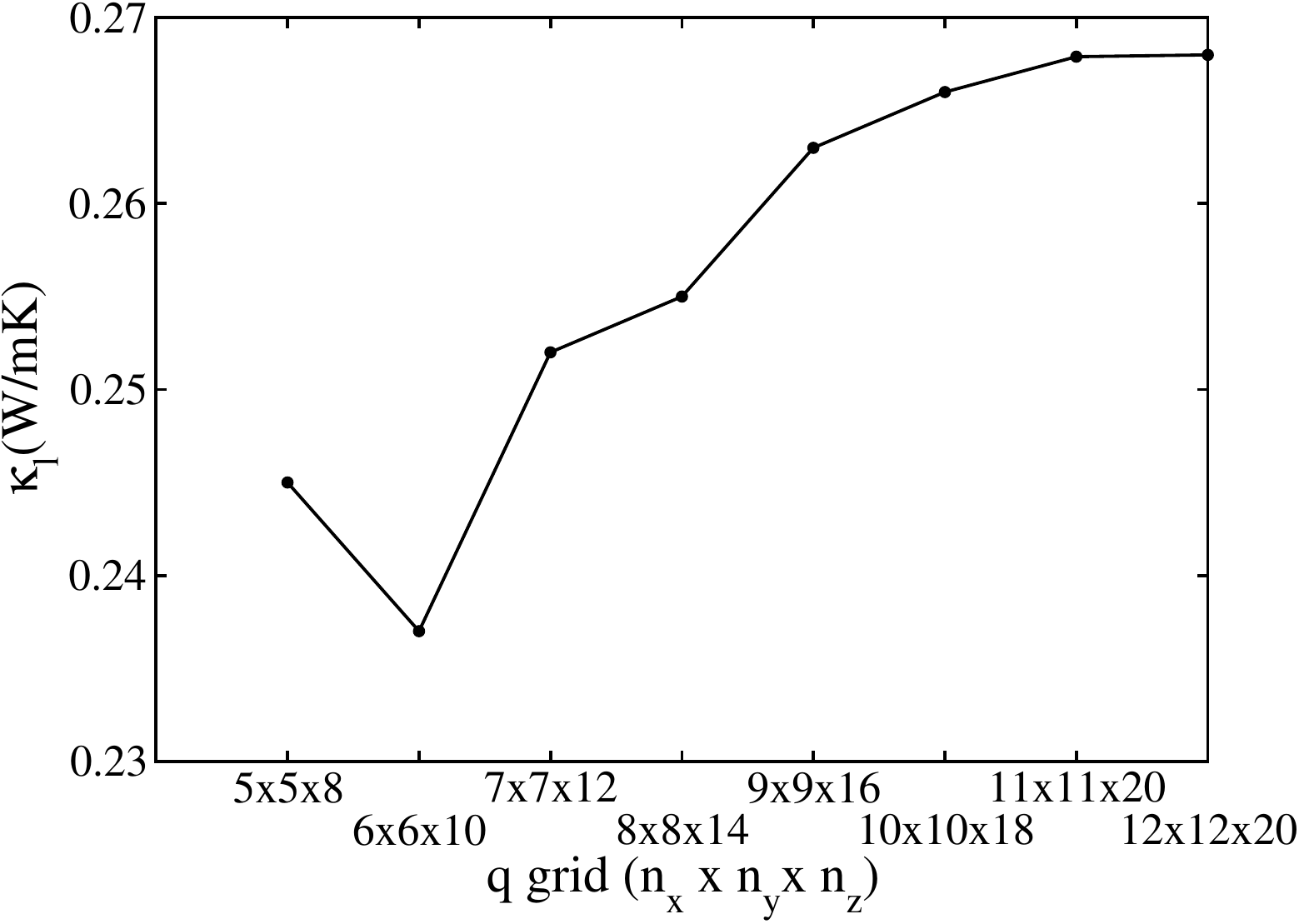}
% \caption{\mathrm{Figure of merit (zT) of (a) p-type (b) n-type of Cs$_2$AgPdCl$_5$ and (c) p-type (d) n-type of Cs$_2$AgPtCl$_5$}  }
%\label{pdzt}
\end{figure}
Figure SI3: Convergence of lattice thermal conductivity with respect to q-grid ($n_x \times n_y \times n_z$)
%\item \textbf{Figure S4:} AIMD total energy at 800 K for 2 ps for (a) Cs$_2$AgPdCl$_5$ and (b) Cs$_2$AgPtCl$_5$.
\begin{figure}[H]
 \centering
   \includegraphics[height=10cm]{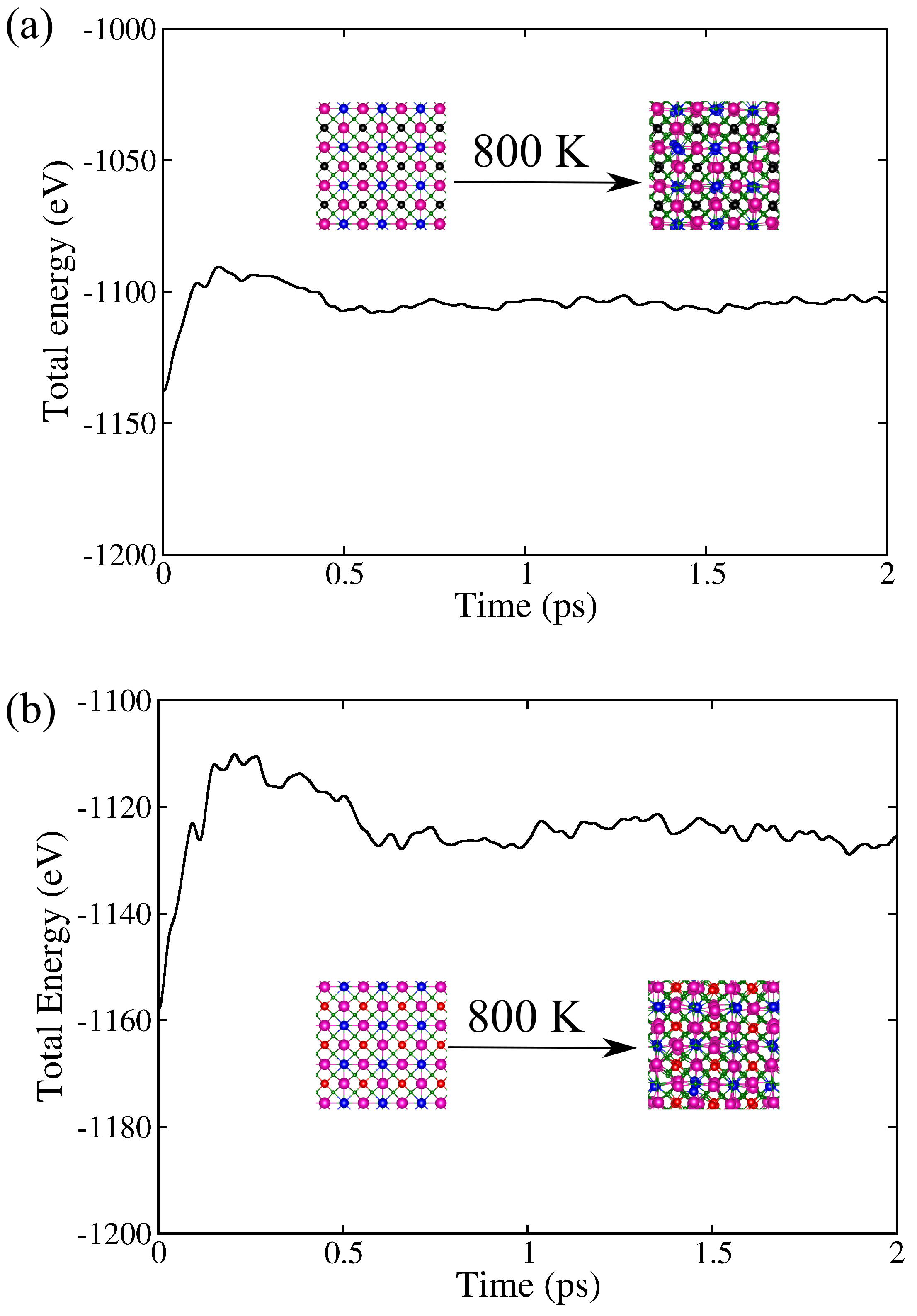}
% \caption{\mathrm{Figure of merit (zT) of (a) p-type (b) n-type of Cs$_2$AgPdCl$_5$ and (c) p-type (d) n-type of Cs$_2$AgPtCl$_5$}  }
%\label{pdzt}
\end{figure}
Figure SI4: AIMD total energy at 800 K for 2 ps for (a) Cs$_2$AgPdCl$_5$ and (b) Cs$_2$AgPtCl$_5$
%\item \textbf{Figure S5:} Projected COHP for (a) Cs$_2$AgPdCl$_5$, (b) Cs$_2$AgPtCl$_5$ for full energy range.
\begin{figure}[H]
 \centering
   \includegraphics[height=8cm]{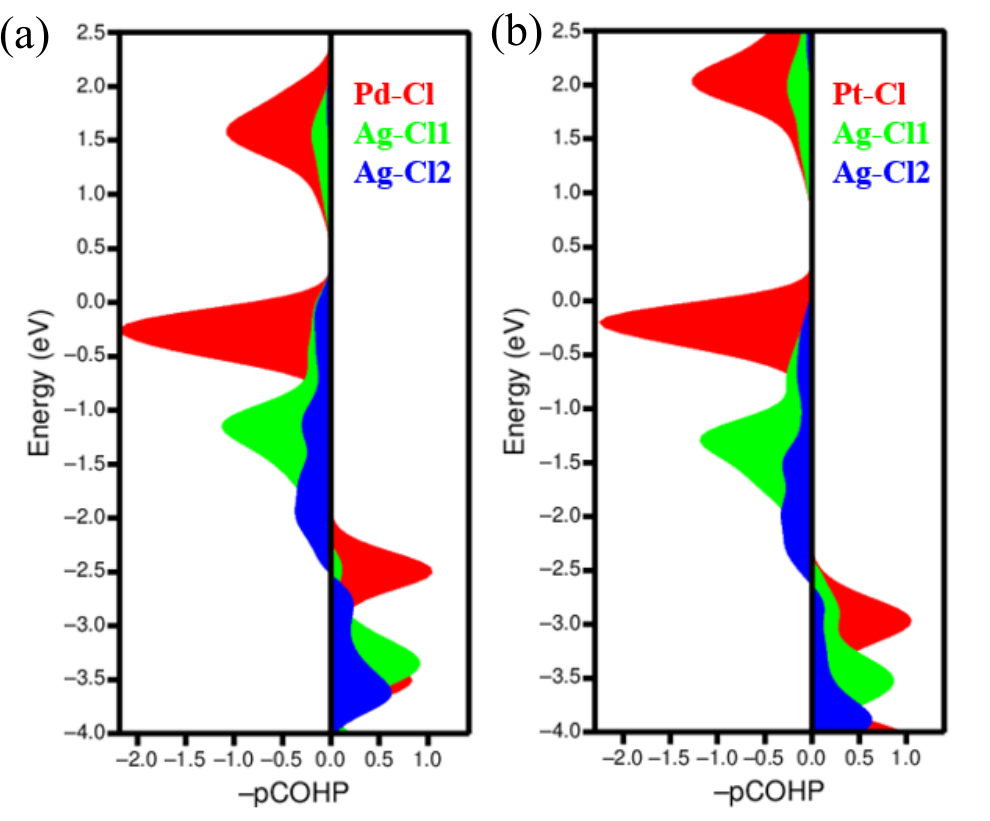}
% \caption{\mathrm{Figure of merit (zT) of (a) p-type (b) n-type of Cs$_2$AgPdCl$_5$ and (c) p-type (d) n-type of Cs$_2$AgPtCl$_5$}  }
%\label{pdzt}
\end{figure}
Figure SI5: Projected COHP for (a) Cs$_2$AgPdCl$_5$, (b) Cs$_2$AgPtCl$_5$ for full energy range
%\item \textbf{Figure S6:} Charge density plots for (a) Cs$_2$AgPdCl$_5$ and (b) Cs$_2$AgPtCl$_5$
\begin{figure}[H]
 \centering
   \includegraphics[height=10cm]{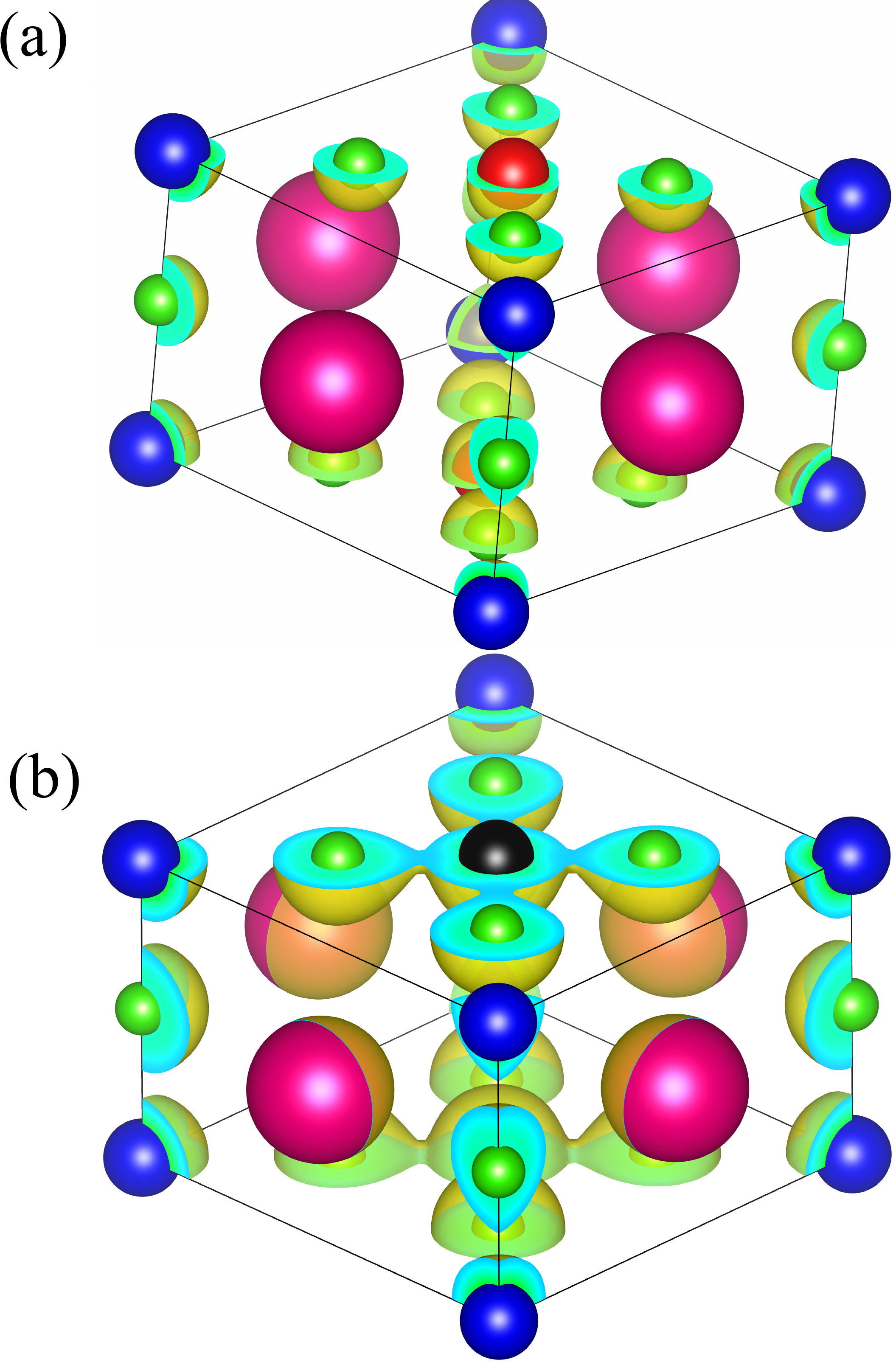}
% \caption{\mathrm{Figure of merit (zT) of (a) p-type (b) n-type of Cs$_2$AgPdCl$_5$ and (c) p-type (d) n-type of Cs$_2$AgPtCl$_5$}  }
%\label{pdzt}
\end{figure}
Figure SI6: Charge density plots for (a) Cs$_2$AgPdCl$_5$ and (b) Cs$_2$AgPtCl$_5$
%\item \textbf{Figure S7:} Electronic band structure along with partial DOS at GGA-PBE + SOC level for (a) Cs$_2$AgPdCl$_5$ and (b) Cs$_2$AgPtCl$_5$.
\begin{figure}[H]
 \centering
   \includegraphics[height=8cm]{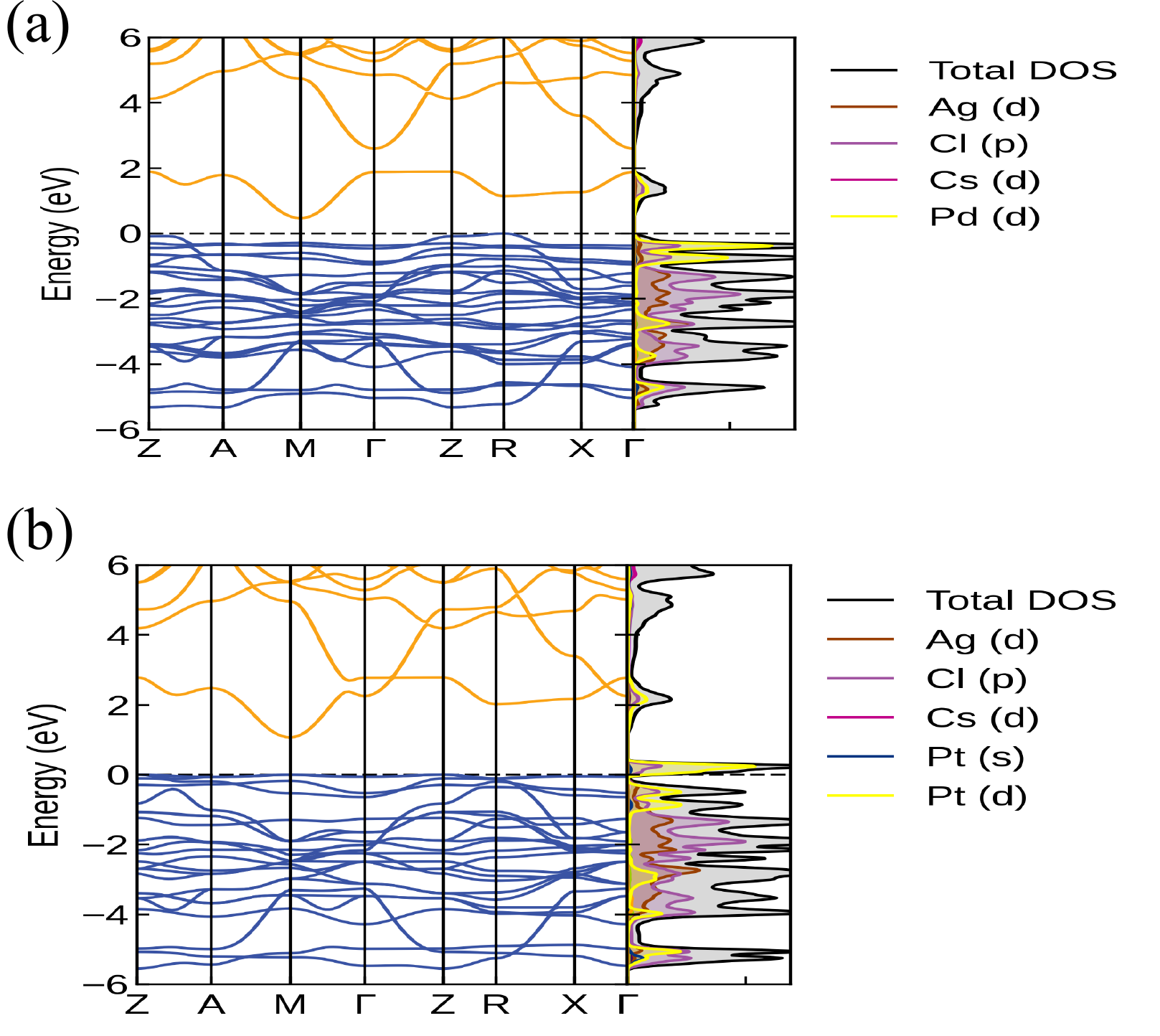}
\end{figure}
Figure SI7: Electronic band structure along with partial DOS at GGA-PBE + SOC level for (a) Cs$_2$AgPdCl$_5$ and (b) Cs$_2$AgPtCl$_5$
%\item \textbf{Figure S8:} Phonon band dispersion and partial DOS without LO-TO splitting for (a) Cs$_2$AgPdCl$_5$ and (b) Cs$_2$AgPtCl$_5$ .
\begin{figure}[H]
 \centering
   \includegraphics[height=8cm]{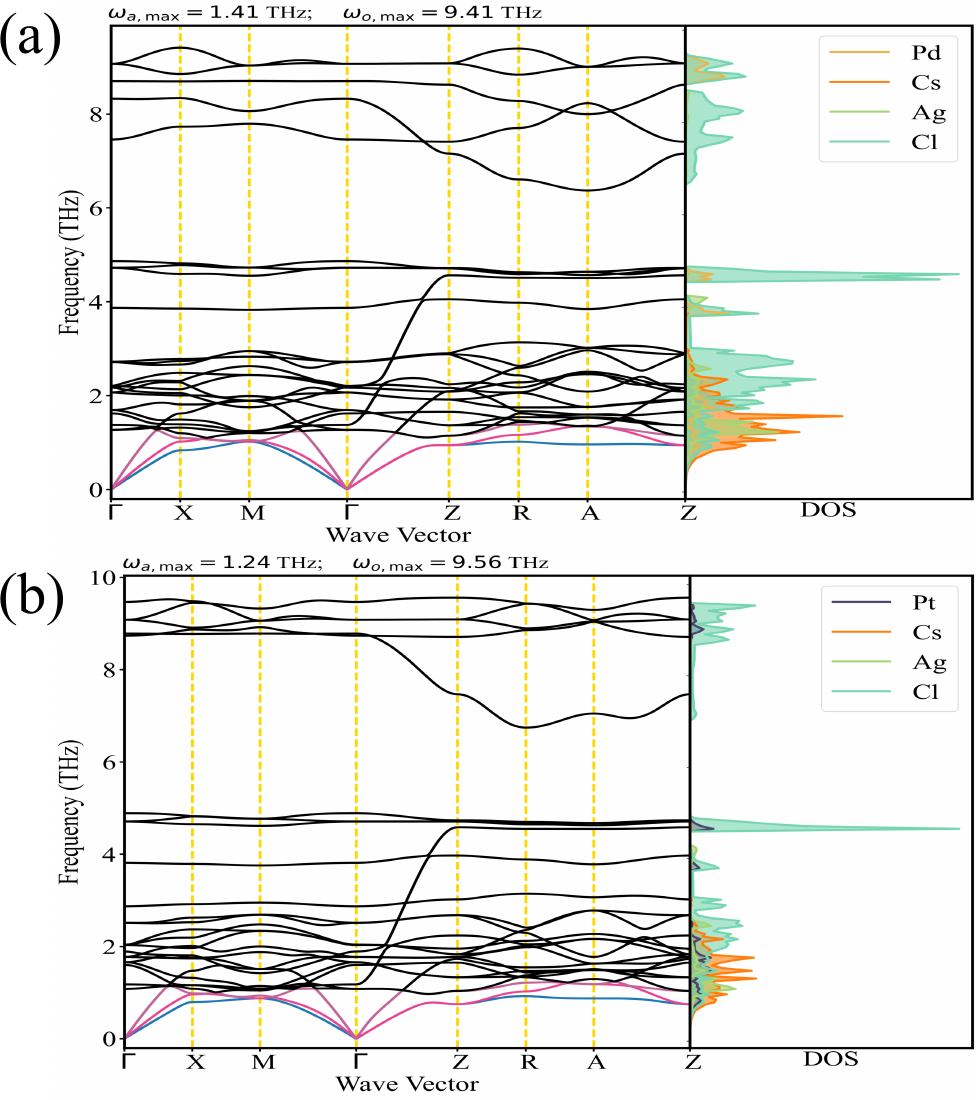}
\end{figure}
Figure SI8: Phonon band dispersion and partial DOS without LO-TO splitting for (a) Cs$_2$AgPdCl$_5$ and (b) Cs$_2$AgPtCl$_5$
%\item \textbf{Figure S9:} Average gr$\ddot{\mathrm u}$neisen parameter as function of temperature for (a) Cs$_2$AgPdCl$_5$ and (b) Cs$_2$AgPtCl$_5$.
\begin{figure}[H]
 \centering
   \includegraphics[height=7cm]{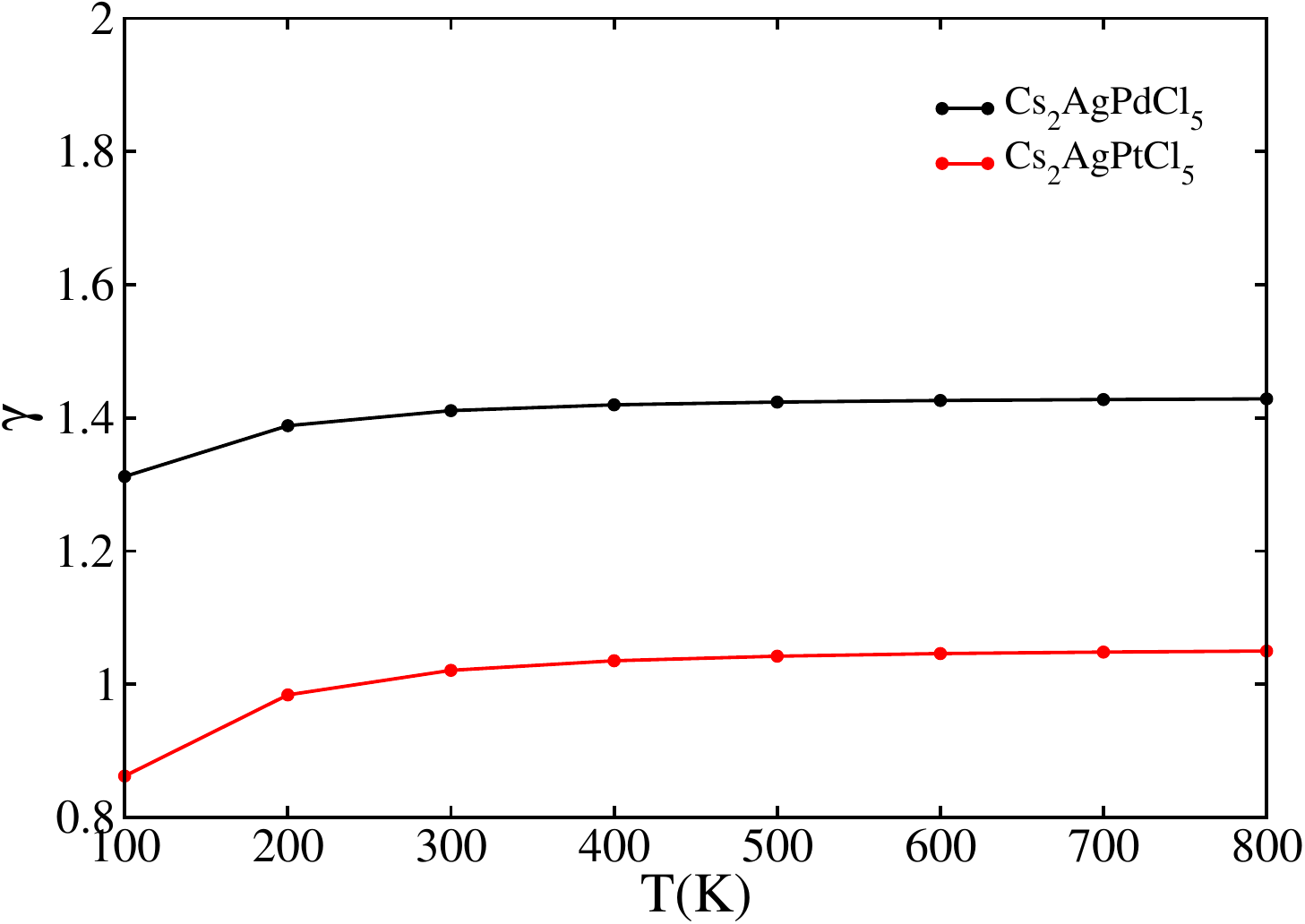}
\end{figure}
Figure SI9: Average gr$\ddot{\mathrm u}$neisen parameter as function of temperature for (a) Cs$_2$AgPdCl$_5$ and (b) Cs$_2$AgPtCl$_5$
%\item \textbf{Figure S10:} Electronic transport as function of carrier concentration and temperature along c-direction for p-type and n-type doping in Cs$_2$AgPdCl$_5$.
\begin{figure}[H]
 \centering
   \includegraphics[height=8cm]{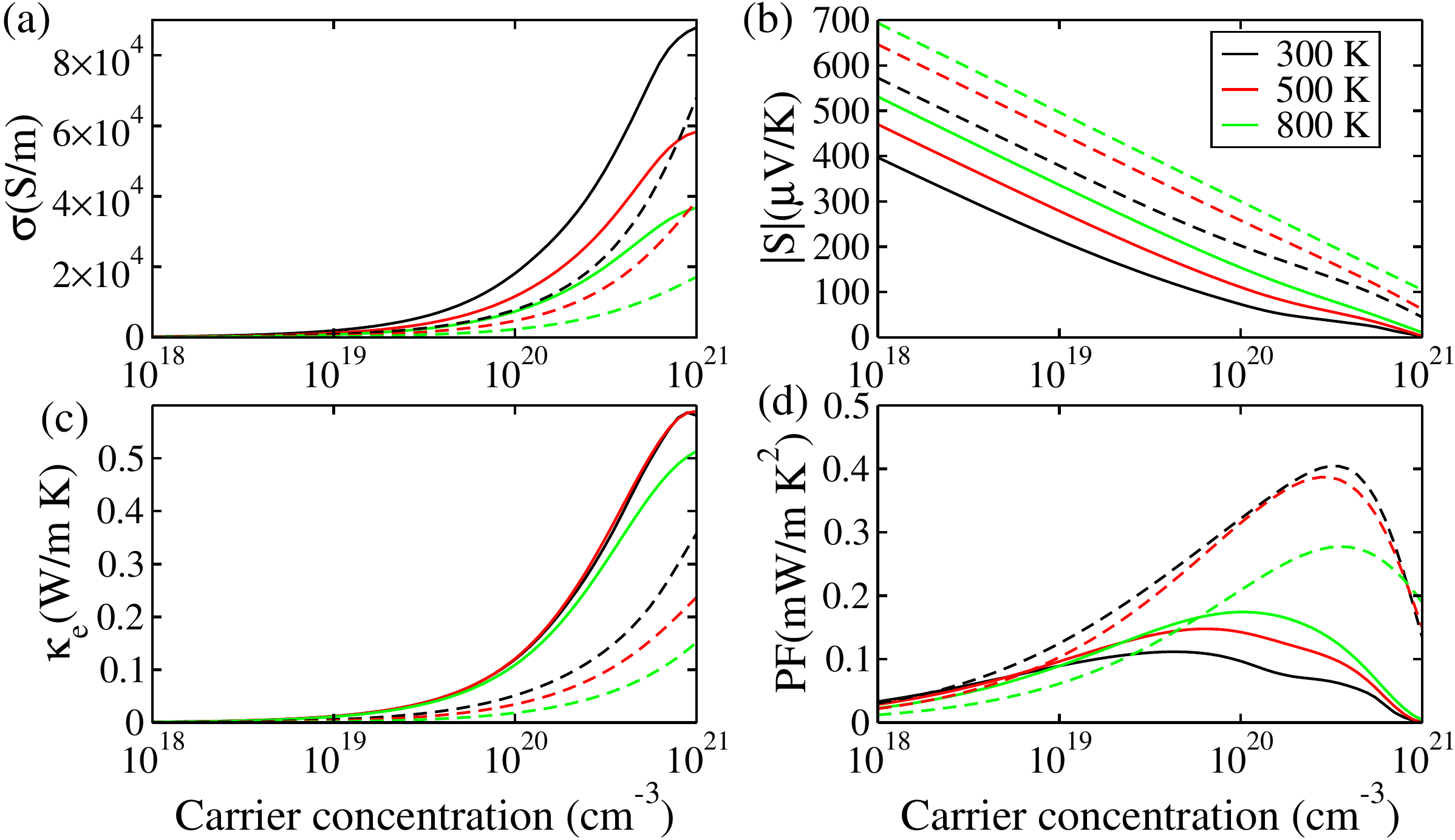}
\end{figure}
Figure SI10: Electronic transport as function of carrier concentration and temperature along c-direction for p-type and n-type doping in Cs$_2$AgPdCl$_5$
%\item \textbf{Figure S11} Electronic transport as function of carrier concentration and temperature along c-direction for p-type and n-type doping in Cs$_2$AgPtCl$_5$.
\begin{figure}[H]
 \centering
   \includegraphics[height=8cm]{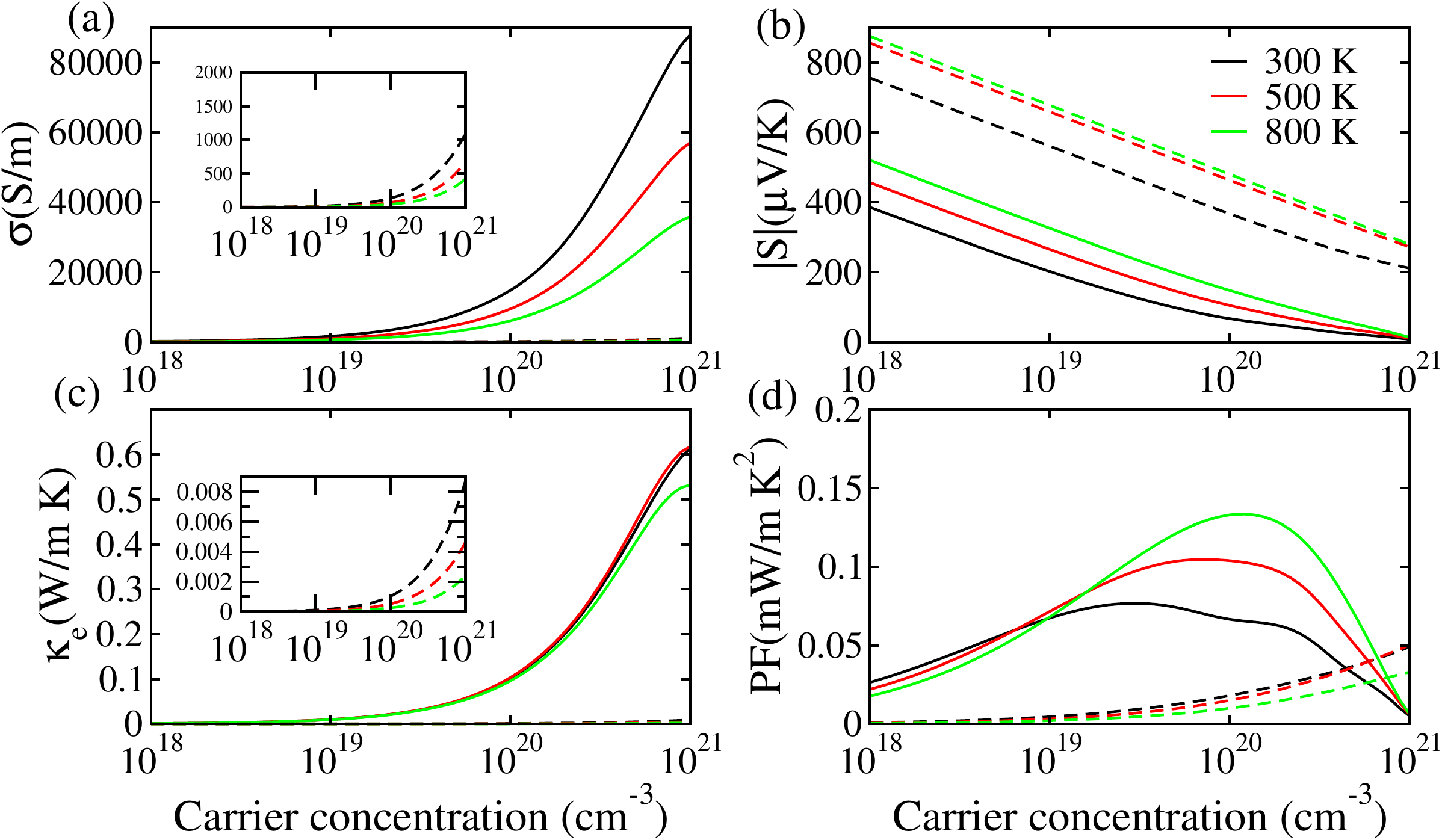}
\end{figure}
Figure SI11: Electronic transport as function of carrier concentration and temperature along c-direction for p-type and n-type doping in Cs$_2$AgPtCl$_5$
%\item \textbf{Figure S12:} Lattice thermal conductivity vs temperature along c-direction (out-of-plane direction).
\begin{figure}[H]
 \centering
   \includegraphics[height=7cm]{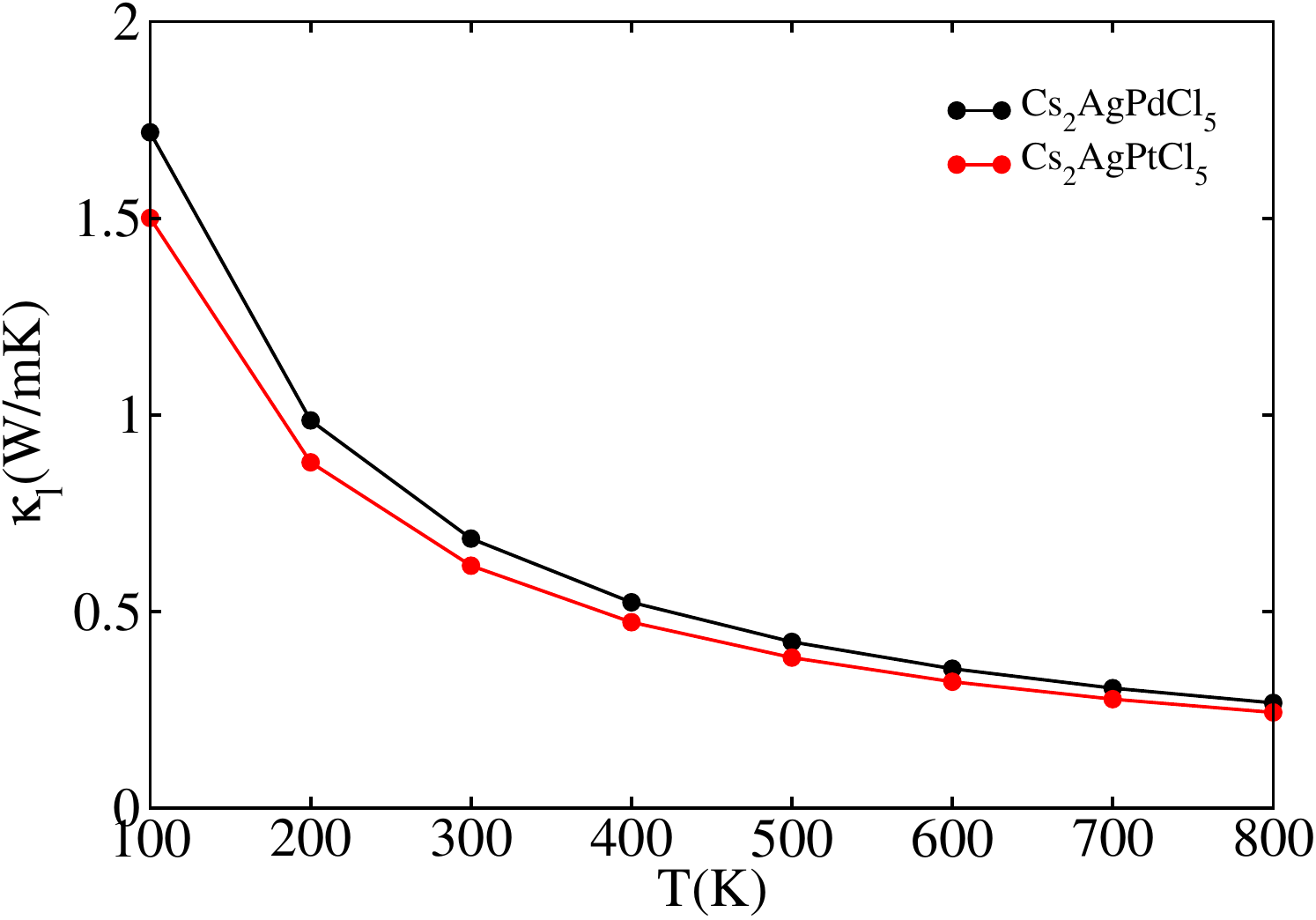}
\end{figure}
Figure SI12: Lattice thermal conductivity vs temperature along c-direction (out-of-plane direction)
%\item \textbf{Figure S13:} Figure of merit (ZT) for p-type and n-type doping in (a) Cs$_2$AgPdCl$_5$ and (b) Cs$_2$AgPtCl$_5$.
\begin{figure}[H]
 \centering
   \includegraphics[height=8cm]{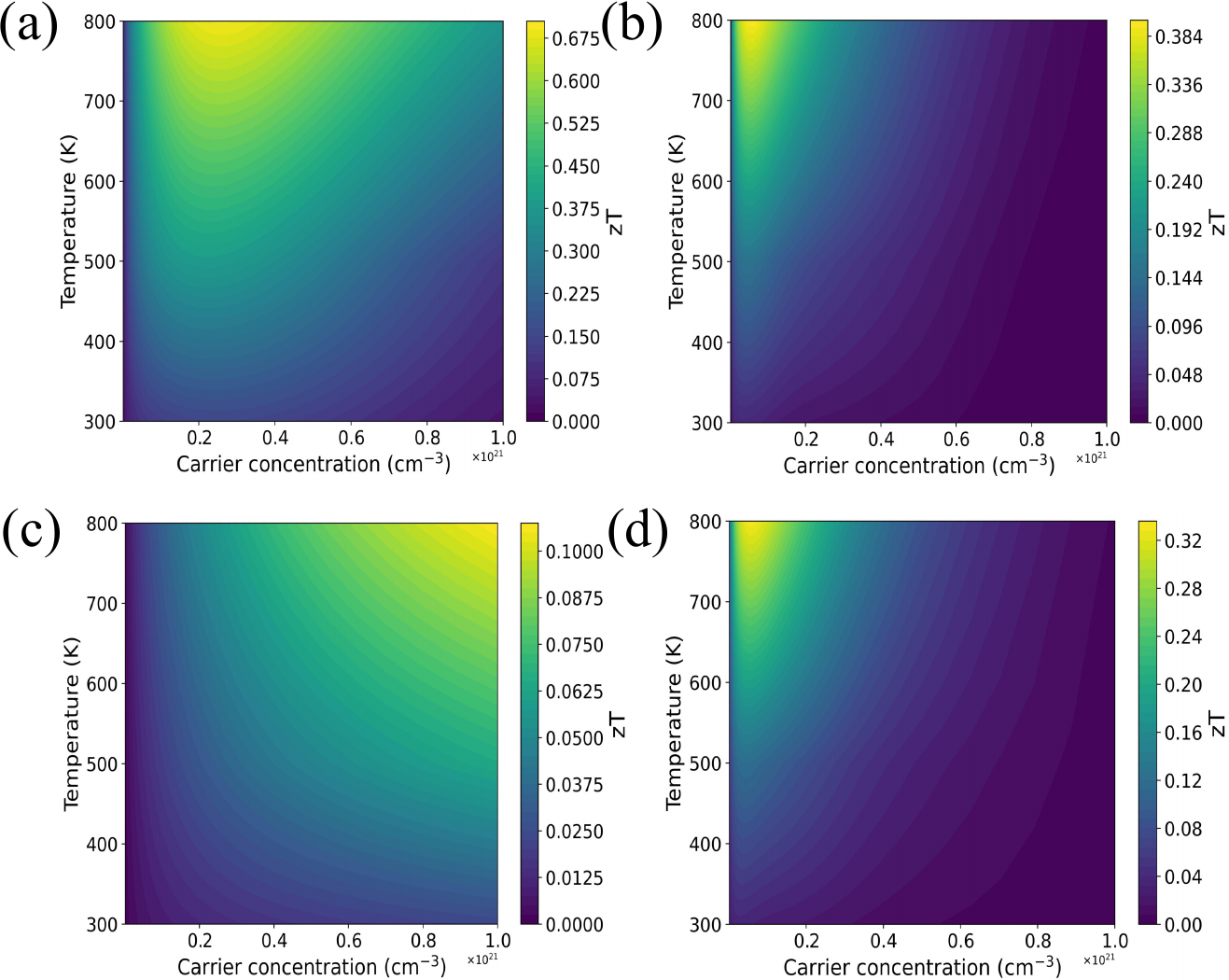}
\end{figure}
Figure SI13: Figure of merit (ZT) for p-type and n-type doping in (a) Cs$_2$AgPdCl$_5$ and (b) Cs$_2$AgPtCl$_5$

%\end{itemize}

%------------------------------------------------
%\sectio*{Supplementary Tables}
%\begin{table}[h!]
\iffalse
\noindent
\begin{center}
\centering
\text{Table S1: Lattice thermal conductivity (W m$^{-1}$ K$^{-1}$) along in-plane direction}
\begin{tabular}{cccccc}
\toprule
T (K) & \multicolumn{2}{c}{$\kappa_l$ (3rd order)} & \multicolumn{2}{c}{$\kappa_l$ (4th order)} \\
\cmidrule(r){2-3} \cmidrule(r){4-5}
 & Cs$_2$AgPdCl$_5$ & Cs$_2$AgPtCl$_5$ & Cs$_2$AgPdCl$_5$ & Cs$_2$AgPtCl$_5$ \\
\midrule
300 & 0.27 & 0.20 & 0.04 & 0.03 \\
500 & 0.17 & 0.13 & 0.02 & 0.01 \\
800 & 0.11 & 0.09 & $\sim$0.01 & $\sim$0.01 \\
\bottomrule
\end{tabular}
\end{center}

\begin{center}
\centering
\text{Table S2: Lattice thermal conductivity (W m$^{-1}$ K$^{-1}$) along c-direction}
\begin{tabular}{cccccc}
\toprule
T (K) & \multicolumn{2}{c}{$\kappa_l$ (3rd order)} & \multicolumn{2}{c}{$\kappa_l$ (4th order)} \\
\cmidrule(r){2-3} \cmidrule(r){4-5}
 & Cs$_2$AgPdCl$_5$ & Cs$_2$AgPtCl$_5$ & Cs$_2$AgPdCl$_5$ & Cs$_2$AgPtCl$_5$ \\
\midrule
300 & 0.69 & 0.62 & 0.15 & 0.14 \\
500 & 0.42 & 0.38 & 0.07 & 0.06 \\
800 & 0.27 & 0.24 & 0.03 & 0.03 \\
\bottomrule
\end{tabular}
\end{center}
\fi

%%%%%%%%%%%%%%%%%%%%%%%%%%%%%%%%%%%%%%%%%%%%%%%%%%%%%%%%%%%%%%%%%%%%%
%% If you are using classical BibTeX rather than biblatex,
%% remove the \printbibliography and uncomment the \bibliograpy one
%%%%%%%%%%%%%%%%%%%%%%%%%%%%%%%%%%%%%%%%%%%%%%%%%%%%%%%%%%%%%%%%%%%%%

\end{document}